%% file: Version2.tex
\pdfoutput=1
\documentclass[12pt,fleqn]{amsart}
\usepackage{amsmath,bm}
\usepackage{amsthm}

\usepackage[colorlinks,hyperindex]{hyperref}%ermöglicht Hyperlinks zwischen Textstellen und zu externen Dokumenten

\usepackage{amsfonts}
\usepackage{latexsym}
\usepackage{amssymb}

\usepackage{graphicx}

\include{MyShortcuts}
\begin{document}
%\frontmatter
\title[Validation of credit default probabilities]{Validation of credit default probabilities via multiple testing procedures}
\author{Sebastian D\"ohler\\
University of Applied Sciences Darmstadt
}
%\\
%Darmstadt University of Applied Sciences}
\address{Sebastian D\"ohler\\ University of Applied Sciences\\
Faculty of Mathematics and Science\\
D-64295 Darmstadt\\
Germany}
\email{sebastian.doehler@h-da.de}
\date{\today}
   % AMS keywords (used in AMS journals)
\begin{abstract}
We apply multiple testing procedures to the validation of estimated default probabilities in credit rating systems. The goal is to identify rating classes for which the probability of default is estimated inaccurately, while still maintaining a predefined level of committing type I errors as measured by the familywise error rate (FWER) and the false discovery rate (FDR). For FWER, we also consider procedures that take possible discreteness of the data resp. test statistics into account. The performance of these methods is illustrated in a simulation setting and for empirical default data.
\end{abstract}
\maketitle
%\mainmatter
\input{Introduction}
\input{Notation}
\input{MTPDiscrete}
\input{SimulationStudy}
%\input{ExampPF}
%\input{ApplicMTP}
%\input{CompHL}
\input{StandardPoors}
\input{Extension}
\input{Discussion}
\begin{appendix}
\include{Tables}
%\newpage
%\pagebreak
\include{Figures}
\end{appendix}
\bibliographystyle{plain}
\label{bibliography}
\bibliography{Vorlage}
%\clearpage
%\part*{Appendix}
%\appendix
%\pagebreak
%\newpage
%\part*{Appendix}
\end{document}

%% file: MyShortcuts.tex
\usepackage{url}		% Für urls
\usepackage{xspace}		% Korrekte Leerzeichen nach Makros
\usepackage{booktabs}

\usepackage{rotating}	% For rotated graphics
\usepackage{fancyvrb}% Für R-Code (5.4.2009)

\newtheorem{prop}{Proposition}

\newtheoremstyle{break}% name
  {9pt}%      Space above, empty = `usual value'
  {9pt}%      Space below
  {}% Body font
  {}%         Indent amount (empty = no indent, \parindent = para indent)
  {\bfseries}% Thm head font
  {}%        Punctuation after thm head
  {\newline}% Space after thm head: \newline = linebreak
  {}%         Thm head spec

\graphicspath{{../..//}}

\newcommand{\erw}{\textnormal{E}}

\newcommand{\NormVert}{\textbf{{N}}}
\newcommand{\BinVert}{\textbf{{Bin}}}
\newcommand{\MinP}{\textnormal{Min}\,$P$\xspace}

\newcommand{\FDR}{\textnormal{FDR}}
\newcommand{\FWER}{\textnormal{FWER}}

%% file: Introduction.tex
\section{Introduction}
Banks use rating systems to classify borrowers according to their credit risk. These systems form the basis for pricing credits and for determining risk premiums and capital requirements for the bank, cf. \cite{Tasche05}. One of the key components in this set-up is the probability of default (PD), i.e. the likelihood with which borrowers will default in a prespecified time period. Banks that use an internal ratings-based (IRB) approach as described in the Basel II framework, are required to report a PD estimate for each borrower. In practice, borrowers are grouped together into rating grades for which a pooled or average PD is calculated. Correct calibration of a rating system means that the respective PD estimates or forecasts are accurate. Inaccurate PD forecasts can lead to substantial losses, see \cite{Stein05} and \cite{BL06}. Since correct calibration is crucial to the appropriate functioning of a rating system, banks are also required by regulatory authorities to validate their PD estimates by comparing the forecasted PD to realized default rates (\cite[\S463 and \S464]{Basle203}). This process is also known as backtesting. Validation of PD estimates can be carried out simultaneously for all rating grades in a joint statistical test or separately for each rating grade, depending on whether an overall assessment or an in detail examination is intended, cf. \cite{Tasche05}. In this paper we are primarily concerned with the latter aim.

The goals of backtesting can vary and will depend on the viewpoint of the involved parties. If backtesting is performed with the aim of demonstrating calibration to a regulatory authority, the bank could be interested in controlling the probability that a correctly calibrated rating system is dismissed or has to be recalibrated. If the goal is to provide more detailed internal information for a bank's risk management it may be desirable to also consider more liberal or exploratory methods that generate early warnings that can help to identify and subsequently investigate potential shortcomings of the rating system.

From a statistical viewpoint, PD validation could be described as the simultaneous assessment of the predictive quality of multiple probability forecasts. In practice, the main statistical methods used for PD validation are (exact or asymptotic) binomial and chi-square tests as well as the so called 'normal test' and various 'traffic light approaches' (cf. \cite{Tasche05} and \cite{BlochMartWehn06} for more details). The binomial and normal tests as well as the traffic light approaches are applied separately to each rating grade whereas the chi-square or Hosmer-Lemeshow test is a global test that can asses several rating categories simultaneously. The normal test and the more exploratory traffic light approaches are multi-period tests which are based on normal approximations and can take dependencies into account. Bl\"ochlinger and Leippold (\cite{BloeLeipp10}) develop a new global goodness-of-fit test for probability forecasts and apply this to empirical default data. Their test consists of two components and they show that the corresponding test statistic is asymptotically $\chi^2$-distributed.

There are several statistical issues associated with PD validation, of which we only mention the two major ones (for more details see e.g. \cite{Tasche05} and \cite{BlochMartWehn06}). Firstly, default data is usually sparse and sample sizes are often small. Combined with PD estimates that usually are very small numbers, this means that the respective hypothesis tests possess low power. Moreover, default events are generally not independent and therefore PD estimates and validation methods should take this into account. Since the main purpose of this paper is to introduce some new concepts to the problem of per-class PD validation we assume for the sake of simplicity and to limit the scope of this paper that default events are independent. However, we give some indication of how the methods described here can be adapted to correlated default data in section \ref{sec:ExtensionDependentDefaults}.

Our aim is to describe some statistical tools that can be used to provide in detail assessments of rating systems, specifically we are concerned with identifying conspicuous resp. miscalibrated rating classes. Suppose a rating system consists of 20 rating grades and for each grade a test is performed at the 5\% significance level. If all null hypotheses are true, then the expected number of false rejection, i.e. erroneously detected miscalibrations will be one. If in addition, the respective test statistics are independent, then the probability of observing at least one false miscalibration finding will be $1-(1-0.05)^{20} \approx 0.64$, i.e. the probability of committing at least one type I error is far from being controlled at the 5\% level. This and related phenomena is known as the multiple testing problem or the problem of multiple comparisons, see \cite{LehmannRomano}. For PD validation, this means that even if all rating classes are perfectly calibrated, the chances of observing a significant finding resp. erroneously concluding that some classes were miscalibrated, is quite high. The problem therefore is to decide which of the significant classes can still be considered miscalibrated and which were identified merely due to performing a multiple number of tests. To the best of our knowledge, this problem has received little attention within the context of PD validation. Rauhmeier (\cite{Rauhmeier06}) takes the problem partly into account in the construction of an original test, which is based on the sum of the number of gradewise rejections. However, as this test is designed as an overall goodness-of-fit test, it can not identify single conspicuous PD estimates. Huschens (\cite{Huschens04}) considers several simultaneous tests and mentions that the Bonferroni procedure (see section \ref{ssec:Bonferroni-type methods}) is inappropriate due to its conservacy. Since his considerations take place in an asymptotic setting however, he also emphasizes that these tests may produce inacceptable results for rating classes with sparse data. In fact, he poses the question how a simultaneous testing procedure could be developed that takes into account the sparseness of data in some rating classes and data richness in others. We attempt to give an answer to this question in section \ref{ssec:Comments}.

Multiple testing procedures (in the sequel abbreviated as MTPs) provide a well-established methodological framework for dealing with multiplicity issues, with several monographs (cf. e.g. \cite{HochbergTamhane1987}, \cite{WestYoung93} and \cite{Hsu1996}) and a large number of research papers available. While MTPs have been used in many areas of application such as clinical trials, microarray experiments, astronomy and magnetic resonance imaging, the validation of PD (and more generally probability) forecasts constitutes to the best of our knowledge a novel field of application.

The plan for this paper is as follows. In section \ref{sec:NotationAssumptions} some further background is given on PD validation and the associated testing problems. Section \ref{sec:ReviewMTP} reviews some multiple testing procedures with a focus on discrete distributions. These procedures are applied in a simulation study in section \ref{sec:SimStudy} and to empirical data in section \ref{sec:EmpiricalStudy}. Following this, an extension to dependent defaults in a single-factor model is sketched. The paper concludes with a discussion in section \ref{sec:Discussion}.

%% file: Notation.tex
\section{Notation and assumptions} \label{sec:NotationAssumptions}
In this section we introduce some terminology and assumptions that will be used in the sequel.
\subsection{The backtesting approach} 
We consider credit portfolios consisting of a total number $N$ of borrowers who have been classified into $K$ rating classes. Each rating class is associated with a true but unknown (average) PD $p_1,\ldots,p_K$ as well as estimated PDs $pd_1, \ldots, pd_K$. The basic idea of the backtesting approach is to split the total sample into a training or estimation sample and a validation sample.
\begin{enumerate}
	\item In the first step, the probability forecast resp. classifier is constructed based on the training sample. In practice, the training sample usually consists of data collected up to some time point $t$ and estimators for the default probability resp. rating classes are usually assigned to individual borrowers based on a vector $x$ of features (covariates) associated with the borrower. Popular models for the dependency of the default probability on $x$ are logistic and probit regression but also nonlinear methods like neural networks and decision trees are used in this context, cf. e.g. \cite{HayPor06}. Note that in this paper we are not concerned with the construction of PD forecasts resp. classifiers but only with assessing the accuracy of a given forecast. Therefore we assume in the sequel that this probability forecast has already been constructed.
	\item The validation sample usually consists of data observed during some future time period, e.g. between $t$ and $t+1$. We denote by $\widehat{n}_j$ the number of borrowers that were assigned to probability forecast $pd_j$ resp. rating class $j$ (say at time $t$) and let $o_j$ denote the number of defaults observed in the rating class between $t$ and $t+1$. Then the true probabilities of default can be estimated e.g. by the quantities $o_1/\widehat{n}_1, \ldots o_K/\widehat{n}_K$ and the quality of the probability forecast resp. classifier can be assessed by statistical tests as described in the introduction.  
\end{enumerate}
\subsection{Testing calibration hypotheses} \label{ssec:TestCalibHyp}
For given $\widehat{n}_1, \ldots, \widehat{n}_K$ and $o_1, \ldots, o_K$ it is to be decided, whether the probability forecasts $pd_1, \ldots, pd_K$ are correct. For $j=1, \ldots, K$ and $l=1,\ldots, n_j$ let $X_{lj} \in \{0,1\}$ denote the rv that indicates whether borrower $l$ in rating grade $j$ defaults ($X_{lj}=1$) or not ($X_{lj}=0$). We assume throughout this paper that 
\begin{tabbing}
(\textbf{A}) \quad \=  $X_{lj} \sim \BinVert(1,p_j)$ and all $X_{lj}$ are independent.
\end{tabbing}
As mentioned in the introduction, independence between all default events is an unrealistic assumption. However, our primary goal is to describe some general MTP approaches to the calibration of PD forecasts. For the clarity of exposition and to concentrate on the main concepts we therefore defer dealing with dependency issues to future work (see also section \ref{sec:ExtensionDependentDefaults}). 
In case of perfect probability forecasts we would have $O_j \sim \BinVert(\widehat{n}_j,p_j)$, where 
\begin{align*}
O_j &= X_{1j}+ \cdots +X_{\widehat{n}_j j}
\end{align*}
and we define null hypotheses accordingly in this probability model as
\begin{align}
H_0^j&: p_j =pd_j && \text{vs.} && H_1^j: p_j \neq pd_j, \label{eq:DefCalHypotheses}
\end{align}
and we say that rating class $j$ is calibrated correctly if $H_0^j$ holds true. In the same spirit we call the probability forecast calibrated in the overall sense if the global hypothesis
\begin{align*}
H_0&:=H_0^1 \cap \cdots \cap H_0^K \tag{P}
\end{align*}
holds true, i.e. if it is calibrated for all rating classes. Note that we consider throughout this paper two-sided hypotheses only. This can be interpreted as the viewpoint of the bank's risk manager who is interested in detecting both overly optimistic and overly pessimistic PD estimates, while regulatory authorities may focus only on one-sided tests that detect underestimation of PDs. However, the MTP approach introduced in section \ref{sec:ReviewMTP} can straightforwardly be adapted to the one-sided case. 

For the simulation experiments in section \ref{sec:SimStudy} it will be helpful to view the problem of forecasting PDs as a classification problem. Suppose that it is known that the true possible default probabilities are given by $p_1,\ldots,p_K$. In this case, the problem of PD forecasting becomes one of PD classification, i.e. deciding for each borrower which of the $p_j$ is true. We denote by $n_1, \ldots, n_K$ the true number of borrowers in classes $1, \ldots, K$. Ideal forecasting resp. perfect classification would mean that $\widehat{n}_j=n_j$ for $j=1,\ldots,K$. In reality we will usually encounter a certain amount of misclassification. To describe this we introduce
\begin{align*}
\widehat{n}_{ij} := \#  &\text{borrowers (truly) from class $i$ that are classified as belonging to class $j$.} 
\intertext{Therefore}
\widehat{n}_{j} &= \widehat{n}_{1j} + \cdots + \widehat{n}_{Kj}
\intertext{and it follows from assumption (A) that the distribution of $O_j$ is given by a convolution of binomial distributions:}
O_j &\sim \overset{K}{\underset{i=1}{\ast}} \BinVert (\widehat{n}_{ij},p_i).
\end{align*}
The matrix $\widehat{N} = (\widehat{n}_{ij})_{1\le i,j \le K}$ is also known as the misclassification or confusion matrix (see e.g. \cite{JohnsonWichern}) and the expectations and variances of the absolute default frequencies can be expressed conveniently through the elements of $\widehat{N}$ and the given PDs.

%The expectations and variances of the absolute default frequencies can be expressed conveniently through the elements of $\widehat{N}$ and the given PDs.
%\begin{prop}
%\begin{itemize}
%	\item[(a)] Generally it holds 
%	\begin{align*}
%\erw_j &:= \erw(O_j) = \sum_{i=1}^{K }\widehat{n}_{ij} \cdot p_i;\\
%V_j&:= \var(O_j) = \sum_{i=1}^{K } \widehat{n}_{ij} \cdot p_i (1- p_i).
%\end{align*}
%\item[(b)] Under $H_0$ it holds
%	\begin{align*}
%\erw_j^{0} &:= \erw(O_j) = \widehat{n}_{j} \cdot p_j;\\
%V_j^{0}&:= \var (O_j) = \widehat{n}_{j} \cdot p_j (1- p_j).
%\end{align*}
%\end{itemize}
%\end{prop}
%\subsection{Statistical tests}\label{ssec:StatTests}
For testing the grade-wise calibration hypotheses $H_0^1,\ldots,H_0^K$ we use an exact binomial test, see also comment (ii) in section \ref{ssec:The Min(P)-approach for discrete distributions}. For testing the global hypothesis $H_0$ in the setting introduced above, a $\chi^2$ goodness-of-fit test based on the statistic
\begin{align*}
T_{HL} &:= \sum_{j=1}^K \frac{(O_j - E_j^0)^2}{V_j^0}
\end{align*}
is commonly used, where $E_j^0=\widehat{n}_{j} \cdot p_j$ resp. $V_j^0=\widehat{n}_{j} \cdot p_j (1- p_j)$ denote the expectation resp. variance of $O_j$ under $H_0$. Hosmer and Lemeshow used a related statistic for assessing the fit of logistic regression models. In \cite{HosmerLemeshow2000} they discuss two methods of grouping risks based on ranked probability estimates:
\begin{enumerate}
	\item In the 'deciles of risk' approach, groups of equal numbers of risks are formed.
	\item In the 'fixed cutpoint' approach, risks are mapped into classes determined by predefined probability cutpoints on the $(0,1)$ interval. This is essentially the approach usually taken in PD validation.
\end{enumerate}
Under appropriate asymptotic conditions (e.g. all $E_j^0$ should be sufficiently large), $T_{HL }$ is approximately $\chi^2$-distributed under $H_0$. It has been demonstrated in \cite{HosmerLemKlar1988} that the deciles of risk approach yields a better approximation to the corresponding $\chi^2$ distribution than the fixed cutpoint approach, especially when the estimated probabilities are smaller than e.g. 0.2. For more details on the advantages and disadvantages of Hosmer-Lemeshow type tests, see \cite{HosmerEtAl1997}. When the sample size is too small to justify the use of asymptotic methods (as is often the case for credit portfolios), the distribution of $T_{HL}$ under $H_0$ can be determined by simulation, cf. \cite{Rauhmeier06}. The corresponding test can be seen as an exact version of the HL-test which corrects for the finite sample size and is denoted by (HL) in the sequel.
%\begin{prop}\footnote{Beibehalten? Wo ander hin?} Let the ... (notation?) from proposition ... hold. Then it holds
%\begin{itemize}
%	\item[(a)] 
%	\begin{align}
%T_{HL} &= \sum_{j=1}^K \frac{V_j}{V_j^0} \cdot \left( \frac{O_j-E_j}{\sqrt{V_j}} +\frac{E_j-E_j^0}{\sqrt{V_j}}      \right)^2 .
%\end{align}
%\item[(b)] If $(O_j-E_j)/\sqrt{V_j}\sim \NormVert(0,1)$ iid then
%	\begin{align}
%T_{HL} &\sim \sum_{j=1}^K Y_j^2 \qquad \text{where $Y_j \sim \NormVert(\delta_j,\sigma_j^2)$ with}\\
%\delta_j &=\frac{E_j-E_j^0}{\sqrt{V_j}},\\
%\sigma_j^2 &=\frac{V_j}{V_j^0}.
%\end{align}
%\end{itemize}
%\end{prop}
%The second statement of the proposition means that $T_{HL}$ is distributed as a linear combination of independent non-central $\chi^2$-variables with linear coefficients given by $\sigma_1^2,\ldots, \sigma_K^2$ and non-centrality parameters given by ... Under the (complete) null hypothesis ...\\
%
% 
%

%% file: MTPDiscrete.tex
\section{A review of some multiple testing procedures}\label{sec:ReviewMTP}
In order to limit the scope of this paper, we confine our review to a selection of classical multiple testing procedures as well as the \MinP approach. For a more complete treatment we refer to the literature on multiple testing cited in the introduction.

We are interested in simultaneously testing a family $H^1_0, \ldots, H^K_0$ of null hypotheses while controlling the probability of one or more false rejections at a multiple level $\alpha$. This probability is called the family-wise error rate (FWER). To be more precise, we require strong control of FWER, i.e. that $\FWER \le \alpha $ holds for all possible constellations of true and false hypotheses. The principal application we have in mind in the context of PD validation are hypotheses tests for binomial proportions (cf. section \ref{ssec:TestCalibHyp}). From the validation viewpoint, it seems highly desirable that apart from controlling the FWER, the multiple testing method employed should possess high power in order to detect possible departures from calibration.

In the sequel let $pv_1,\ldots,pv_K$ denote the $p$-values observed from testing hypotheses $H^1_0, \ldots, H^K_0$ and assume that these values are ordered $pv_1 \le \cdots \le pv_K$.

\subsection{Bonferroni-type methods} \label{ssec:Bonferroni-type methods}
The Bonferroni method (in the sequel abbreviated as (Bonf)) is a classical method that maintains control of the FWER. Adjusted $p$-values  are defined by $pv'_j:= \max( K \cdot pv_j,1)$ and all hypotheses with $pv'_j \le \alpha$ are rejected. 

Instead of using the (single-step) Bonferroni method one can use the more powerful Holm step-down (from the most significant to the least significant result) procedure (Hol) which works the following way: Define adjusted $p$-values by
\begin{align*}
pv'_1 &:= K \cdot pv_1,\\
pv'_2 &:= \max(pv'_1, (K-1) \cdot pv_2),\\
pv'_3 &:= \max(pv'_2, (K-2) \cdot pv_3),\\
 &\vdots \\
 pv'_K &:= \max(pv'_{K-1}, pv_K)
\end{align*}
and again set the adjusted $p$-values exceeding 1, to 1. All hypotheses with $pv'_j \le \alpha$ can then be rejected. Another variant of Bonferroni-type adjustment which is more powerful than Holm's procedure is Hommels (Hom) approach which is valid under independence or positive independence assumptions (for details refer to \cite{Hommel1988}).

All the procedures described above provide strong control of the FWER under certain circumstandes. For the Bonferroni and Holm procedure this holds true e.g. when the distribution functions of the $p$-values, considered as random variables $PV_1,\ldots,PV_K$, are stochastically larger under the respective null hypothesis than some uniformly distributed random variable, i.e. for $i=1,\ldots,K$ it holds $P(PV_i \le u | H^j_0) \le u$ for all $u \in (0,1)$, cf.  \cite{LehmannRomano}. However, as noted e.g. in \cite{WestWolf1997}, these procedures can be very conservative, especially if the $p$-values are distributed discretely. Therefore it makes sense to investigate multiple testing procedures developed specifically for discrete distributions.
\subsection{The \MinP approach for discrete distributions} \label{ssec:The Min(P)-approach for discrete distributions}
Gutman and Hochberg review and compare the performance of several FWER controlling MTPs for discrete distributions (cf. \cite{Gutman2007} and the references cited therein). They investigate Tarone's method, two variants of a method by Roth, the method of Hommel and Krummenauer, the \MinP method of Westfall and collaborators (see \cite{WestTroend2008} and the references cited therein) and an original method called $TWW_k$. All methods except the \MinP method and the method of Hommel and Krummenauer lack $\alpha$-consistency. This means that possibly a hypothesis cannot be rejected at some level $\alpha_1$ but can be rejected at some lower level $\alpha_2$. In addition it is shown in \cite{Gutman2007} that the \MinP method is universally more powerful than the method of Hommel and Krummenauer. Since $\alpha$-consistency would seem to be a desirable property in the validation context considered here, we concentrate in the sequel on the more powerful method of the two, namely the \MinP approach.

\subsubsection{The single-step version}
Suppose the distribution of $\min(PV_1,\ldots, PV_K)$, when all null hypotheses are true, is available. For the single-step variant the idea of the \MinP approach is to define adjusted $p$-values by
\begin{align*}
pv'_j &:= P(\min(PV_1,\ldots, PV_K) \le pv_j)
\end{align*} 
where $pv_1, \ldots,pv_K$ are the $p$-values observed for the data, i.e. the $j$th adjusted $p$-value is the probability that the minimum $p$-value is smaller than the $j$th observed $p$-value. In \cite{WestWolf1997} it is pointed out that this quantity measures the 'degree of surprise that the analyst should experience after isolating the smallest $p$-value from a long list of $p$-values calculated from a given data set.' For the relationship of the \MinP procedure to some other MTPs and its use in the analysis of toxicology data, see \cite{WestWolf1997} as well. 
\subsubsection{The step-down version}
Corresponding to the single-step method described above, a step-down variant can be constructed. Following \cite{WestTroend2008} define $H_I=\cap_{i \in I} H^i_0$ for $I \subset \{1,\ldots,K \}$. Suppose again the observed $p$-values are $pv_1 \le \cdots \le pv_K$, corresponding to null hypotheses $H^1_0, \ldots, H^K_0 $ then define adjusted $p$-values
\begin{align}
pv'_j&:= \max_{i\le j} pv_{\{i,\ldots,K\}} \label{eq:MinPadjPV}
\intertext{where $p_I$ is given by}
pv_I &=P(\min_{i \in I} PV_i \le \min_{i \in I} pv_i|H_I). \notag
\end{align}
The decision rule 'reject $H^j_0$ if $pv'_j \le \alpha$' yields a procedure which controls the FWER at level $\alpha$ if the so called 'subset pivotality condition' is fulfilled. Subset pivotality means that the distribution of any subvector of $p$-values under the respective null hypotheses is unaffected by the truth or falsehood of the remaining hypotheses, i.e.
%\begin{tabbing}
% \qquad	\textbf{(SPC)} \qquad \=  \> For all $I \subset \{1,\ldots,K \}$ the distributions of $(PV_i)_{i \in I} | H_I$ and \\
% \>$(PV_i)_{i \in I}  | H_{\{1,\ldots,K\}}$ are identical.
%\end{tabbing}
\begin{tabbing}
 \qquad	\textbf{(SPC)} \qquad \=   For all $I \subset \{1,\ldots,K \}$ the distributions of $(PV_i)_{i \in I} | H_I$ and \\
 \>$(PV_i)_{i \in I}  | H_{\{1,\ldots,K\}}$ are identical.
\end{tabbing}
For the \MinP approach (SPC) implies that the distribution of $\min_{i \in I}PV_i | H_I$ and $\min_{i \in I}PV_i |  H_{\{1,\ldots,K\}}$ are identical, cf. \cite{WestTroend2008} for the relationship of this method with the closure principle in multiple testing. Clearly, (SPC) holds if the distribution of each $PV_j$ depends only on the validity if $H_0^j$.
\begin{prop}\label{prop:SuffCondSPC}
Let $H_0^1, \ldots, H_0^K$ be (general) hypotheses with associated $p$-value rv's $PV_1, \ldots,PV_K$. If the distribution of each $PV_j$ only depends on the validity of $H_0^j$, i.e. for all $j$ and $I \subset \{1,\ldots,K \}$ with $j \in I$ it holds 
\begin{align}
 PV_j|H_I &\sim PV_j|H_0^j \label{eq:SPCmarginal}
\end{align}
 then (SPC) holds true.  
\end{prop}
%\begin{proof}
%Let $I=\{i_1, \ldots, i_m \}$ and define $M_I:= \min_{i \in I} PV_i$. Then it holds 
%\begin{align*}
%M_I|H_I &\sim \min(PV_{i_1}(O_{i_1}), \ldots, PV_{i_m}(O_{i_m}))|H_0^{i_1} \cap \cdots \cap H_0^{i_m}
%\intertext{and since the distribution of $O_j$ does not depend on any $H_l$ with $l \neq j$ we have}
%&\sim \min(PV_{i_1}(O_{i_1})|H_0^{i_1}, \ldots, PV_{i_m}(O_{i_m})|H_0^{i_m})
%\end{align*}
%and therefore $M_I|H_I \sim M_I|H_J$ for all $I \subset J$ and in particular for $J=\{1,\ldots,K \}$.
%\end{proof}
\begin{proof}
It holds that
\begin{align*}
(PV_i)_{i \in I}|H_I &\sim (PV_i|H_I)_{i \in I}\\
&\sim (PV_i|H_0^j)_{i \in I}   \qquad \text{by \eqref{eq:SPCmarginal}}\\
&\sim (PV_i|H_{\{1,\ldots,K\}})_{i \in I}   \qquad \text{by \eqref{eq:SPCmarginal}}\\
&\sim (PV_i)_{i \in I}|H_{\{1,\ldots,K\}}.
\end{align*}
\end{proof}
\paragraph{Comments:}
\begin{itemize}
	\item[(i)] We can apply proposition \ref{prop:SuffCondSPC} to the basic set-up introduced in section \ref{ssec:TestCalibHyp}. Let $PV_j:=PV_j(O_j)$, where $O_j$ denote the number of observed defaults in rating grade $j$. Clearly, for the hypotheses given by \eqref{eq:DefCalHypotheses} condition \eqref{eq:SPCmarginal} then holds true. In the sequel we will use the exact binomial test to calculate these $p$-values, but we note that this is not essential to our approach and that principally any test that controls the type I error on the test-wise level could be used.
	\item[(ii)] As described in \cite{WestWolf1997}, there is some controversy concerning the definition of two-sided $p$-values for discrete tests. Generally, different types of two-sided $p$-values will affect not only the observed $p$-values, but also their distribution and therefore also the \MinP adjusted $p$-values. For the calculation in this paper two-sided $p$-values implemented in the R-function binom.test are used. These values are based on the sum of probabilities of events with smaller likelihood than the observed data, see also \cite{Hirji06}. 
	\item[(iii)] Note that proposition \ref{prop:SuffCondSPC} is also applicable in the case of dependent $p$-values as long as condition \eqref{eq:SPCmarginal} is satisfied. 
	\item[(iv)] In order to calculate the adjusted $p$-values in \eqref{eq:MinPadjPV}, the distribution functions
	\begin{align*}
F_{\{i,\ldots,K\}} (x)&:= P(\min_{l \in \{i,\ldots,K\}} PV_l \le x|H_{\{i,\ldots,K\}})
\end{align*}
have to determined. In general, simulation techniques will have to be used to accomplish this, but for the case where all observations are independent there is a simpler way, which is described in the next section.
\end{itemize}
\subsubsection{Determining the \MinP distribution function for independent $p$-values} \label{sssec:DetermineMinP}
If $PV_1, \ldots, PV_K$ are independent we have 
\begin{align*}
F_{\{1,\ldots,K\}} (x)&= F^{Ind}_{\{1,\ldots,K\}} (x) := 1-\prod_{i=1}^K (1-F_i(x))
\end{align*}
where $F_i(x)=P(PV_i\le x)$ is the distribution function of the $i$th $p$-value under $H_{\{1,\ldots,K\}}$. 
Let $A_i:= \{pv_{it} | t=1,\ldots,m_i \}$ denote the ordered possible values of $PV_i$ under $H_{\{1,\ldots,K\}}$, i.e. $0< pv_{i1}< \cdots <pv_{i m_i}$. When the distribution of $PV_i$ is discrete, then $F_i$ is a (right-continuous) step function with jump discontinuities at abscissa values $pv_{i1}< \cdots <pv_{i m_i}$. If no assumption on the dependency structure of $(PV_1, \ldots, PV_K)$ is made, the Bonferroni inequality yields the following conservative bound
\begin{align*}
F_{\{1,\ldots,K\}} (x)&\le F^{Bonf}_{\{1,\ldots,K\}} (x) = \min(\sum_{i=1}^K F_i(x),1).
\end{align*}
This means that both $F^{Ind}_{\{1,\ldots,K\}}$ and $F^{Bonf}_{\{1,\ldots,K\}}$ are step functions with discontinuities at the values $A= \cup_{i=1}^K A_i$. Assume that the set $A$ of all possible $p$-values in the experiment consists of ordered values $0<x_1<\cdots < x_M$. Now define for $i=1, \ldots, K$ and $l=1,\ldots ,M$
\begin{align*}
y_{il} &:= F_i(x_l).
\intertext{Then it holds} 
y^{Ind}_l &:=1 - \prod_{i=1}^K (1-y_{il}),\\
y^{Bonf}_l &:= \min(\sum_{i=1}^K y_{il},1),
\end{align*}
and the values $y^{Ind}_l $ resp. $y^{Bonf}_l $ are the ordinate values of $F^{Ind}_{\{1,\ldots,K\}}$ resp. $F^{Bonf}_{\{1,\ldots,K\}}$. %For the numerical implementation ist may be convenient\footnote{Bringt das wirklich einen Mehrwert?} to reexpress $y_{il}$ as
%\begin{align}
%y_{il} &= \sum_{\{ j: pv_{ij}\le x_l\}} z_{ij}
%\intertext{where $z_{ij} $ is defined for $j=1, \ldots,m_i$ as}
%z_{ij} &=
%\begin{cases}
%F_i(pv_{i1}) & \qquad \text{for $j=1$}\\
%F_i(pv_{ij})-F_i(pv_{i,j-1}) & \qquad \text{for $j=2, \ldots,m_i$}
%\end{cases}\\
%&=
%\begin{cases}
%pv_{i1} & \qquad \text{for $j=1$}\\
%pv_{ij}-pv_{i,j-1} & \qquad \text{for $j=2, \ldots,m_i$}
%\end{cases}
%\end{align}
%i.e. for fixed $i$ the vector $z_{ij}$ consists of the smallest possible $p$-value, followed by the difference vector.

The approach described above for ${\{1,\ldots,K\}}$ carries over directly to index sets ${\{j,\ldots,K\}}$ and so the $p$-values needed for the determination of the adjusted $p$-values in \eqref{eq:MinPadjPV} can be obtained by
\begin{align*}
pv_{\{i,\ldots,K\}}&=  F_{\{i,\ldots,K\}} (\min(pv_i,\ldots,pv_K))
\end{align*}
where $F_{\{i,\ldots,K\}}$ is $F^{Ind}_{\{i,\ldots,K\}}$ if the $p$-values are independent or could be chosen conservatively as $F^{Bonf}_{\{i,\ldots,K\}}$ in the general dependency case. In the examples considered in section \ref{sec:SimStudy}, the differences between the discrete Bonferroni \MinP method (d-Bonf) and the discrete independence \MinP method (d-Ind) are mostly minimal. Therefore we concentrate in the sequel on (d-Bonf) and the corresponding step-down method (sd-d-Bonf). For dealing with specific forms of dependencies, power can be gained by using the simulation approaches mentioned above.
\paragraph{Example} To compare the \MinP approach with the continuous Bonferroni resp. independence corrections we consider $K=11$ hypotheses given by $H^j_0: O_j \sim \BinVert(n_j,p_j)$ with
\begin{align*}
(n_1, \ldots,n_{11})=&(31,17,7,8,7,6,7,2,5,8,2),\\
(p_1, \ldots,p_{11})=&(0.00015,0.0003,0.00060,0.0011,0.002,0.0035,0.006,\\
& \ 0.0105,0.0185,0.0325,0.057).
\end{align*}
Figure \ref{fig:DistrTMIN} shows the uniform and discrete versions of the distribution functions $F^{Ind}_{\{1,\ldots,11\}}$ and $F^{Bonf}_{\{1,\ldots,11\}}$ where the discrete versions were obtained by the method described above. (For the uniform (continous) case we have $F_i(x)=x$ for $x\in [0,1]$.)
%\begin{figure}[h]
%	\centering
%		%\includegraphics[width=1.00\textwidth]{DistrTMIN.pdf}
%		\includegraphics[height=0.33\textheight]{DistrTMIN.pdf}
%	\caption{Distribution functions of $F^{Ind}_{\{1,\ldots,11\}}$ and $F^{Bonf}_{\{1,\ldots,11\}}$ in the continuous and the discrete case}
%	\label{fig:DistrTMIN}
%\end{figure}
\begin{center}
[Fig. \ref{fig:DistrTMIN} about here]
\end{center}
This figure shows that the difference between the continuous and the discrete approach is considerable, whereas in either case there seems to be no relevant difference between the independence or the slightly more conservative Bonferroni correction. For the usual significance level $\alpha=.05$ the \MinP-based critical value (under independence or Bonferroni dependence) is $c^{\textnormal{Min} P}_{0.05}\approx 0.0139$, i.e. $H_{\{1,\ldots,11\}}$ is rejected if the minimum $p$-value observed in the eleven hypotheses tests is less than or equal to this value. For the continuous approaches we have $c^{Ind}_{0.05}\approx 0.0047$ and $c^{Bonf}_{0.05}\approx 0.0045$. From the viewpoint of the continuous approaches, the \MinP-based critical value therefore corresponds to an effective number of three tests instead of eleven.

As stated in \cite{WestWolf1997}, the benefit of the \MinP approach generally 'depends on the specific characteristics of the discrete distributions. Larger gains are possible when $K$ is large, and where many variables are sparse'. 
\subsection{False discovery rate}\label{sec:FDR}
Instead of controlling the FWER, the algorithm of Benjamini and Hochberg (\cite{BenjaminiHochberg95}) and related methods seek control of the 'false discovery rate' (FDR), where a false discovery occurs whenever a null hypothesis is erroneously rejected. Let $m_0$ denote the (unknown) number of true hypotheses, $V$ the number of true hypotheses that are erroneously rejected by some given MTP, let $R$ be the total number of rejected hypotheses and set $Q:=V/\max(R,1)$. Then the FDR is defined as $\FDR=\erw(Q)$. When all null hypotheses are true, then $\FDR=\FWER$ and when $m_0 \le K $, then $\FDR \le \FWER$, see \cite{BenjaminiHochberg95}. Hence, any procedure that controls FWER also controls FDR, but if only control of FDR is desired, these methods are potentially much more powerful than the methods described in the preceding sections, especially when the number of tests is large. In the context of PD validation they could serve as explorative tools as mentioned in the introduction.

The Benjamini-Hochberg (BH) procedure consists of rejecting $H_0^1, \ldots,H_0^k$ where $k$ is determined by
\begin{align*}
k &= \max \{i | pv_i \le \frac{i}{K} \cdot \alpha \}.
\end{align*}
If no such $i$ exists, no hypothesis is rejected. (FDR-) adjusted $p$-values are defined in step-down fashion (cf. \cite{ReinerYekBenj03}):
\begin{align*}
pv'_K &:= pv_K,\\
pv'_{K-1} &:= \min(pv'_K, \frac{K}{K-1}\cdot pv_{K-1}),\\
pv'_{K-2} &:= \min(pv'_{K-1}, \frac{K}{K-2}\cdot pv_{K-2}),\\
 &\quad \vdots \\
 pv'_{1} &:= \min(pv'_{2}, K \cdot pv_{1}).
\end{align*}
The (BH) procedure then consists of rejecting all hypotheses with $pv'_j \le \alpha$. If the underlying rv's $PV_1, \ldots, PV_K$ are independent then it can be shown that $\FDR\le m_0 \cdot \alpha /K$ holds true, with equality holding if the test statistics are continuous (cf. \cite[Theorem 5.1]{BenjaminiYekutieli01}), and since $m_0 \le K$, the FDR is controlled at level $\alpha$. 

For discrete data, specific modifications of the (BH) procedure have been proposed by Pounds and Chen (cf. \cite{PoundsCheng06}) resp. Gilbert (cf. \cite{Gilbert05}) in the context of analysing gene expression resp. genetics data. The method of Pounds and Chen is derived under the assumption that the proportion $m_0/K$ of true hypotheses to the total number of hypotheses is sufficiently small, i.e. it is assumed that $P((PV_1+\cdots+PV_K)/K>1/2)\approx 0$. Since the number $K$ is a relatively small number in our applications, this appears to be an inappropriate restriction. Gilbert's  modification of (BH) uses Tarone's method which lacks $\alpha$-consistency, as noted in the beginning of this section. For these reasons we have refrained from evaluating these methods in sections \ref{sec:SimStudy} and \ref{sec:EmpiricalStudy}.

The power of the (BH) procedure can be increased by suitably estimating $m_0$ and then incorporating this estimate by applying (BH) to $\alpha':= \alpha \cdot K/ \widehat{m}_0$ instead of $\alpha$, if $\widehat{m}_0>0$. This results in the adaptive BH algorithm, which we denote hereafter by (a-BH). The particular estimator $\widehat{m}_0$ used here is motivated by a graphical approach originally proposed by Schweder and Sp{\o}tvoll (cf. \cite{BenjaminiHochberg00} for more details). Further adaptive FDR procedures which may yield more power are described in \cite{BenjaminiKriegerYekutieli06} but are not investigated here.

Although in this paper we are only concerned with independent $p$-values, we note that there are some results resp. modifications available for dealing with dependent $p$-values. Benjamini and Yekutieli show that under the most general dependency structure the (BH) procedure still controls the FDR at level $\alpha \cdot (1+1/2+1/3+ \cdots + 1/K)$, cf. \cite[Theorem 1.3]{BenjaminiYekutieli01}. In \cite{YekutieliBenjamini99} they also investigate simulation based approaches which allow more specific modelling of dependency structures.

\subsection{Comments} \label{ssec:Comments}
The MTPs introduced in this section provide flexible and versatile tools for the task of PD validation. Conceptually they allow a separation of the validation process into two steps. 
\begin{enumerate}
	\item In the first step, tests for the individual calibration hypotheses are carried out. This is a step which has to be performed in any case due to regulatory requirements. The results of these tests can be presented in terms of $p$-values. As noted in \cite{WestWolf1997}, using $p$-values instead of the original test statistics may be more appropriate when tests are discrete, since $p$-values are on the same scale, whereas test statistics, which are often based on counts, are generally not.
	 
	The only general requirement at this stage is that the employed tests should be as powerful as possible, subject to the control of the type I error. But apart from this, the MTP approach allows liberty in the choice of which specific test is used. For instance, it would also be possible to accomodate traffic light approaches, as long as the results can be expressed in terms of $p$-values.
	\item In the second step, the marginal $p$-values are combined by some appropriate MTP like (Bonf), (d-Bonf) or (BH), yielding multiplicity adjusted $p$-values resp. rejected calibration hypotheses. 
\end{enumerate}
An additional advantage of using approaches based on $p$-values, is that this provides a flexible and natural way of dealing with heterogeneous tests over different classes. In particular, it answers the question of Huschens mentioned in the introduction: If data is sparse in some classes and rich in some others it is possible to use e.g. asymptotic tests for the data rich classes while using exact tests for the others. 

%% file: SimulationStudy.tex
\section{Simulation study}\label{sec:SimStudy}
In this section we introduce a simple prototype credit portfolio and two types of misclassification matrices that will describe incorrect PD calibration. With these components we will assess the performance of MTPs for identifying conspicuous rating classes and for rejecting the global hypothesis. 

For a compact description of the results, we group the MTPs from section \ref{sec:ReviewMTP} in the following way:
\begin{itemize}
	\item group A consists of the Bonferroni-based procedures (Bonf), (Hol) and (Hom),
	\item group B consists of the FDR procedure (BH) and its adaptive modification (a-BH), and
	\item group C consists of the \MinP methods (d-Bonf) and (sd-d-Bonf) (and in some instances also (d-Ind)).
\end{itemize}

All calculations in this paper were done using the statistical software package R, see \cite{RSoft1}. For all grade-wise tests the exact binomial test implemented in the R-function binom.test was used. The (Bonf), (Hol), (Hom) and (BH) adjusted $p$-values were calculated using the R-function p.adjust. For (a-BH) the R-package fdrtool was used for estimating $m_0$. The code for (d-Bonf), (d-Ind) and (sd-d-bonf) was implemented by the author.
\subsection{Example portfolio and misclassification matrices} \label{sec:ExampPFMisClass}
\subsubsection{Example portfolio}
We consider a rating system consisting of $K=14 $ classes. As described in section \ref{sec:NotationAssumptions}, we assume that borrowers are assigned to one of these classes by some given model or mechanism. The credit portfolio consists of given true PDs $p_1, \ldots, p_{14}$ and some probability distribution $q_1, \ldots, q_{14}$ of borrowers to rating classes (cf. table \ref{tab:TableSimStudy}). 
%\begin{itemize}
%	\item Typical (average) probabilities of default $(PD_1,\ldots,PD_{14})=()$ corresponding to the rating classes $1,\ldots,14$.
%	\item A given probability  
%\end{itemize}
%\begin{figure}[h]
%	\centering
%		%\includegraphics[width=1.00\textwidth]{DistrTMIN.pdf}
%		\includegraphics[height=0.33\textheight]{PortfolioGraph.pdf}
%	\caption{Distribution of rating classes for example portfolio (PDs are given below class numbers)}
%	\label{fig:PortfolioGraph}
%\end{figure} 
These components make up an artificial example but nevertheless contain some typical features. The number $K=14$ of classes may seem large, but even $K=25$ classes are not uncommon. The S\&P rating system considered in the empirical study in section \ref{sec:EmpiricalStudy}, for instance, uses 17 rating classes. Another typical feature is the sparsity of data: Most of the default probabilities are rather small and the numbers of borrowers may also be small in several classes. Specifically, it is to be expected that there may be numerous classes where the distribution of test statistics is quite discrete and asymptotic methods may not be valid.

In the simulation experiments we will consider 10 portfolio sizes of $N_{PF}=100, \ldots,1000$. The portfolios are class-wise increasing in the sense that if $n_j(N_{PF})$ denotes the number of borrowers in the portfolio with true pd $p_j$ we have $n_j(100)\le n_j(200)\le \cdots \le n_j(1000)$ for $j=1, \ldots, 14$ and the relative frequencies of borrowers within the rating classes is roughly equal to $q_1, \ldots, q_{14}$. For each $N_{PF}$ we only draw one realisation of the portfolio, i.e. we ignore the sampling variability that arises from drawing finite sample sizes of borrowers from the distribution. 
\subsubsection{Two types of misclassification schemes}
We introduce two simple types of models for the misclassification matrix $\widehat{N}$ from section \ref{sec:NotationAssumptions}. 
\paragraph{Upgrade-downgrade model}
The upgrade-downgrade model $\widehat{N}^1=\widehat{N}^1(s)$ is parametrized by a shift parameter $s \in \{-K,-K+1, \ldots, K-1,K \}$. Each borrower is upgraded by $s$, i.e. if $g$ is the true rating grade of a borrower he/she will be classified to group
\begin{align*}
\widehat{g} &=
\begin{cases}
1 & \qquad \text{if $g-s \le 0$,}\\
g-s & \qquad \text{if $g-s \in \{1\ldots,K \}$,}\\
K & \qquad \text{if $g-s \ge K+1$.}
\end{cases}
\intertext{resp. for the estimated PD's it holds}
pd_{i} &=
\begin{cases}
p_1 & \qquad \text{if $i-s \le 0$,}\\
p_{i-s} & \qquad \text{if $i-s \in \{1\ldots,K \}$,}\\
p_K & \qquad \text{if $i-s \ge K+1$.}
\end{cases}
\end{align*}
This model respresents a systematic, monotone error in the rating system, resp. PD estimation (cf. \cite{Rauhmeier06}).
If $s>0$ each borrower is rated too optimistically (low rating classes corresponding to low default probabilities) resp. for $s<0$ too pessimistically. For $s=0$ the ideal classification resp. estimation is obtained. 
\paragraph{Example}The matrix below gives an example for $N_{PF}=300$ and $s=-3$. 
\setcounter{MaxMatrixCols}{20}
\begin{align}
\widehat{N}^1(-3) &=
\begin{pmatrix}
$ 2$&$ 0$&$ 0$&$ 0$&$ 0$&$ 0$&$ 0$&$ 0$&$ 0$&$0$&$0$&$0$&$0$&$0$\\
$ 5$&$ 0$&$ 0$&$ 0$&$ 0$&$ 0$&$ 0$&$ 0$&$ 0$&$0$&$0$&$0$&$0$&$0$\\
$14$&$ 0$&$ 0$&$ 0$&$ 0$&$ 0$&$ 0$&$ 0$&$ 0$&$0$&$0$&$0$&$0$&$0$\\
$22$&$ 0$&$ 0$&$ 0$&$ 0$&$ 0$&$ 0$&$ 0$&$ 0$&$0$&$0$&$0$&$0$&$0$\\
$ 0$&$46$&$ 0$&$ 0$&$ 0$&$ 0$&$ 0$&$ 0$&$ 0$&$0$&$0$&$0$&$0$&$0$\\
$ 0$&$ 0$&$39$&$ 0$&$ 0$&$ 0$&$ 0$&$ 0$&$ 0$&$0$&$0$&$0$&$0$&$0$\\
$ 0$&$ 0$&$ 0$&$39$&$ 0$&$ 0$&$ 0$&$ 0$&$ 0$&$0$&$0$&$0$&$0$&$0$\\
$ 0$&$ 0$&$ 0$&$ 0$&$43$&$ 0$&$ 0$&$ 0$&$ 0$&$0$&$0$&$0$&$0$&$0$\\
$ 0$&$ 0$&$ 0$&$ 0$&$ 0$&$32$&$ 0$&$ 0$&$ 0$&$0$&$0$&$0$&$0$&$0$\\
$ 0$&$ 0$&$ 0$&$ 0$&$ 0$&$ 0$&$26$&$ 0$&$ 0$&$0$&$0$&$0$&$0$&$0$\\
$ 0$&$ 0$&$ 0$&$ 0$&$ 0$&$ 0$&$ 0$&$14$&$ 0$&$0$&$0$&$0$&$0$&$0$\\
$ 0$&$ 0$&$ 0$&$ 0$&$ 0$&$ 0$&$ 0$&$ 0$&$16$&$0$&$0$&$0$&$0$&$0$\\
$ 0$&$ 0$&$ 0$&$ 0$&$ 0$&$ 0$&$ 0$&$ 0$&$ 0$&$0$&$0$&$0$&$0$&$0$\\
$ 0$&$ 0$&$ 0$&$ 0$&$ 0$&$ 0$&$ 0$&$ 0$&$ 0$&$0$&$2$&$0$&$0$&$0$\\
\end{pmatrix}
\label{eq:BsplUpDownMatrix}
\end{align}
In this case the 5, 14 and 22 borrowers from rating classes 2, 3 and 4 are upgraded to rating class 1, 46 borrowers from grade 5 are assigned to class 2 etc.

Even though $s$ is a metric variable, it may be more appropriate to interpret its influence on $\widehat{N}^1$ in an ordinal way, i.e. $\widehat{N}^1(2)$ is more pessimistic than $\widehat{N}^1(1)$ but not twice as pessimistic.
\paragraph{Dispersion model}
The other scenario we consider is the model $\widehat{N}^2=\widehat{N}^2(h)$ with dispersion parameter $h \ge 0$ where
\begin{align*}
\widehat{n}^2_{ij}&=\widehat{n}^2_{ij}(h)= [n_i \cdot w_{ij}(h)]
\intertext{where $[\cdot]$ denotes rounding and the wheights are defined by}
w_{ij}(h) &= \frac{W_h(|i-j|)}{\sum_{l=1}^K W_h(|i-l|)}
\intertext{and $W_h$ is defined by the function}
W_h(x) &=\varphi(x/h),
\end{align*} 
where $\varphi$ denotes the density function of $\NormVert(0,1)$. This means that the matrix $W_h$ converges for $h \rightarrow 0$ against the identity matrix, the respective $\widehat{N}^2$ representing the ideal classifier, and in the worst case, for $h \rightarrow \infty$, the number of true borrowers per rating class is dispersed roughly uniformly over all rating classes. Due to rounding differences, the total number of borrowers may change (moderately) for different values of $h$. This model represents a random error in the sense that as $h$ increases, the classification becomes increasingly imprecise. As in the case of $\widehat{N}^1$, this model is only intended as a simple way of obtaining a certain kind of misclassification.

\subsection{Identification of conspicuous rating classes}
We now apply the MTPs introduced in section \ref{sec:ReviewMTP} to the problem of identifying conspicuous rating classes, i.e. rejecting single hypotheses $H^j_0$. In most cases, groups A, B and C show quite distinct behavior.
\subsubsection{Numerical example for a single sample of defaults}
We begin by describing the way the discretized \MinP methods work for a concrete numerical example as given by table \ref{tab:NumExampMTP}. Suppose we have $N_{PF}=300$ borrowers and the misclassification is given by the matrix $\widehat{N}^1(-3)$ from \eqref{eq:BsplUpDownMatrix} in the example above. The entries of $\widehat{N}^1(-3)$ together with the mapping of rating classes to default probabilities yield the first three rows of table \ref{tab:NumExampMTP}. The 300 borrowers have been classified into 10 out of 14 possible classes.  Suppose the observed validation sample is given by row 4 of this table, resulting in $p$-values for the exact (two-sided) binomial test in row 5. The rest of the table consists of the adjusted $p$-values produced by the various multiple testing procedures described in section \ref{sec:ReviewMTP}. 
\begin{table}[htbp]
\centering
{\footnotesize
 \begin{tabular}{lrrrrrrrrrr}\hline\hline
 class $j$ & 1&2 &3 & 4&5 & 6& 7&8 &9 & 11\\
$\widehat{n}_j$ &$43$&$46$&$39$&$39$&$43$&$32$&$26$&$14$&$16$&$2$\\
$pd_j$ &$ 0.0001$&$ 0.0003$&$ 0.0006$&$ 0.0011$&$ 0.0020$&$ 0.0035$&$ 0.0060$&$ 0.0105$&$ 0.0185$&$0.0570$\\
defaults &$ 0$&$ 1$&$ 0$&$ 1$&$ 0$&$ 1$&$ 1$&$ 2$&$ 1$&$1$\\
$p$-values&$ 1.0000$&$ 0.0137$&$ 1.0000$&$ 0.0420$&$ 1.0000$&$ 0.1061$&$ 0.1448$&$ 0.0092$&$ 0.2583$&$0.1108$\\\hline
(Bonf)&$ 1.0000$&$ 0.1371$&$ 1.0000$&$ 0.4202$&$ 1.0000$&$ 1.0000$&$ 1.0000$&$ 0.0923$&$ 1.0000$&$1.0000$\\
(Hol)&$1.0000$&$ 0.1234$&$ 1.0000$&$ 0.3361$&$ 1.0000$&$ 0.7429$&$ 0.7429$&$ 0.0923$&$ 1.0000$&$0.7429$\\
(Hom)&$ 1.0000$&$ 0.1234$&$ 1.0000$&$ 0.2941$&$ 1.0000$&$ 0.5307$&$ 0.6457$&$ 0.0830$&$ 1.0000$&$0.5538$\\\hline
(BH)&$ 1.0000$&$ 0.0685$&$ 1.0000$&$ 0.1401$&$ 1.0000$&$ 0.2215$&$ 0.2414$&$ 0.0685$&$ 0.3690$&$0.2215$\\
(a-BH)&$ 1.0000$&$ 0.0685$&$ 1.0000$&$ 0.1401$&$ 1.0000$&$ 0.2215$&$ 0.2414$&$ 0.0685$&$ 0.3690$&$0.2215$\\\hline
(d-Ind)&$ 1.0000$&$ 0.0551$&$ 1.0000$&$ 0.1428$&$ 1.0000$&$ 0.2906$&$ 0.5237$&$ \textbf{0.0322}$&$ 0.6341$&$0.3671$\\
(d-Bonf)&$1.0000$&$ 0.0564$&$1.0000$&$ 0.1512$&$1.0000$&$ 0.3316$&$ 0.7015$&$ \textbf{0.0327}$&$ 0.9251$&$0.4391$\\
(sd-d-Bonf)&$ 1.0000$&$ \textbf{0.0472}$&$ 1.0000$&$ 0.1291$&$ 1.0000$&$ 0.2666$&$ 0.2915$&$ \textbf{0.0327}$&$ 0.3703$&$0.2680$\\\hline
\end{tabular}}
\caption{Adjusted $p$-values for a single realization of defaults with $N=300$ and upgrade-downgrade alternative with $s=-3$ (significant findings in boldface)}
\label{tab:NumExampMTP}
\end{table} 
Within group A it holds that (Hom) is more powerful than (Hol) which is more powerful than (Bonf), which is a known general result (cf. \cite{LehmannRomano}). However, even (Hom) does not reject any of the hypotheses. For both methods in B identical results hold, i.e. $m_0$ was estimated as $K$. The adjusted $p$-values are substantially lower than for group A but still fail to achieve significance. Within group C no relevant difference between (d-Bonf) and (d-Ind) is discernible but both procedures are outperformed by the step-down version (sd-d-Bonf). These procedures are able to identify one resp. two invalid PD estimates. 

The workings of (d-Bonf) and (sd-d-Bonf) are illustrated in figure \ref{fig:ExampMTPstepdown} for the two smallest $p$-values $pv_8$ and $pv_2$ (represented by ticks on the $x$-axis). 
\begin{center}
[Fig. \ref{fig:ExampMTPstepdown} about here]
\end{center}
%\begin{figure}[bhtp]
%	\centering
%		%\includegraphics[width=1.00\textwidth]{ExampMTPstepdown.pdf}
%		\includegraphics[height=0.33\textheight]{ExampMTPstepdown.pdf}
%	\label{fig:ExampMTPstepdown}
%	\caption{Distribution functions for \MinP for the first and second step in (sd-d-Bonf)}
%\end{figure} 
For (d-Bonf) the distribution function $F_{\{1,\ldots,9,11\}}$, represented by the solid line is determined by the method described in section \ref{sssec:DetermineMinP}. Obviously, it holds $F_{\{1,\ldots,9,11\}}(pv_8)\le 0.05$ but $F_{\{1,\ldots,9,11\}}(pv_2)> 0.05$, so this procedure only rejects $H_0^8$. The procedure (sd-d-Bonf) starts with $F_{\{1,\ldots,9,11\}}(pv_8)$ as well, thereby rejecting $H_0^8$. In the second step, $F_{\{1,\ldots,7,9,11\}}$, represented by the dashed line, is determined which now yields a barely significant result for $pv_2$. In the successive steps, functions $F_{\{1,3,\ldots,7,9,11\}},F_{\{1,3,5,\ldots,7,9,11\}}, \ldots $ are determined, resulting in values $pv_{\{1,3,\ldots,7,9,11\}},pv_{\{1,3,5,\ldots,7,9,11\}}, \ldots $ and the adjusted $p$-values defined by \eqref{eq:MinPadjPV} are listed in the last row of table \ref{tab:NumExampMTP}.
\subsubsection{Simulation results for a single portfolio and misclassification matrix}\label{sssec:SimSinglePF}
For the observed defaults in table \ref{tab:NumExampMTP} the highest number of invalid rating classes were identified by (sd-d-Bonf), the second most by (d-Bonf) and fewer by all other procedures. It would be interesting to see if this picture is due to the specific observation or is representative of the general situation. In order to investigate this, 10000 observations with the true default probabilities were simulated and for each simulation the testing procedures were evaluated as in table \ref{tab:NumExampMTP}. A summary of these results is given in figure \ref{fig:Identify_300_-3_100}.
\begin{center}
[Fig. \ref{fig:Identify_300_-3_100} about here]
\end{center}
%\begin{figure}[htbp]
%	\centering
%		%\includegraphics[width=1.00\textwidth]{DistrTMIN.pdf}
%		\includegraphics[height=0.33\textheight]{Identify_300_-3_10000.pdf}
%	\caption{Simulated probabilities of rejecting null hypotheses corresponding to the rating classes for $N=300$ and upgrade-downgrade alternative with $s=-3$}
%	\label{fig:Identify_300_-3_100}
%\end{figure} 

For the ability to identify validation violations, the findings from table \ref{tab:NumExampMTP} still basically hold true, i.e.:
\begin{itemize}
	\item the procedures from group A possess the lowest power,
	\item group B outperforms group A
	\item group C outperforms group B except for classes 2 and 3, where (BH) and (a-BH) are better than (d-Bonf) and (d-Ind)
	\item the (sd-d-Bonf) procedure is superior to all other procedures.
\end{itemize}
Note also that the main classes identified as questionable are classes with high ratings, i.e. relatively high default probabilities. For the low rating classes the relatively large sample sizes are not able to compensate for the loss of power resulting from the extremely low default probabilities. 

Another measure for comparing the relative power of the procedures is given by the average number of rejections. For the simulations underlying figure \ref{fig:Identify_300_-3_100} the results are listed in table \ref{tab:AveNumRejection_300_-3_100}. 
Again, the result is consistent with previous analyses: group A constitutes the least powerful, group C the most powerful methods. Within the latter group (d-Bonf) and (d-Ind) perform similarly, (sd-d-Bonf) performs best, albeit with only a slight advantage. An intermediate position is taken by group B. 
\begin{table}[hbtp]
	\centering
		\begin{tabular}{cccccccc}\hline\hline
	(Bonf)     &       (Hol)   &       (Hom)   &           (BH)  &     (a-BH) &  (d-Bonf)& (d-Ind) &     (sd-d-Bonf) \\
		0.5625        &  0.5715    &      0.5836      &    0.7589     &    0.7614 & 0.9962   &       0.9962     &     1.0309\\	\hline
		\end{tabular}
		\caption{Average number of rejections for $N_{PF}=300$ and upgrade-downgrade alternative with $s=-3$}
		\label{tab:AveNumRejection_300_-3_100}
\end{table}
\subsubsection{Simulation results for average number of rejections}
\paragraph{Upgrade-downgrade misclassification}
In table \ref{tab:AveNumRejection_300_-3_100} the average number of rejected hypotheses was given for a specific portfolio size and a specific shift value in the upgrade-downgrade model. Figure  \ref{fig:Identify_Ave_num_rejections_UpDownGrade} illustrates corresponding simulation results for varying portfolio sizes $100,\ldots, 1000$ and shifts $s=-5, \ldots,5$. 
\begin{center}
[Fig. \ref{fig:Identify_Ave_num_rejections_UpDownGrade} about here]
\end{center}
%\begin{figure}[hbtp]
%	\centering
%		\includegraphics[width=1\textwidth]{Identify_Ave_num_rejections_UpDownGrade.pdf}
%		%\includegraphics[height=0.33\textheight]{Identify_Ave_num_rejections_UpDownGrade.pdf}
%	\caption{Simulated probabilities of rejecting null hypotheses corresponding to the rating classes for $N=300$ and upgrade-downgrade alternative with $s=-3$ $(N_{Sim}=1000)$}
%	\label{fig:Identify_Ave_num_rejections_UpDownGrade}
%\end{figure} 
Again, the results within the groups are in line with previous analyses. For portfolio sizes up to 400, group C outperforms group B, while for larger portfolios the situation is more ambivalent. For negative shifts, i.e. pessimistic ratings, the best procedure in C appears to be somewhat superior to the best procedure in B and vice versa for positive shifts.

\paragraph{Dispersed misclassification} For this type of alternative the results are similar to those of the upgrade-downgrade alternative. As illustrated in figure \ref{fig:Identify_Ave_num_rejections_Dispersion}, group A is uniformly outperformed by groups B and C. For portfolio sizes up to 600, group C outperforms group B, while for bigger portfolios the procedures in B are superior to C, especially for large values of the dispersion parameter.
\begin{center}
[Fig. \ref{fig:Identify_Ave_num_rejections_Dispersion} about here]
\end{center}

\subsection{Tests for the global calibration hypothesis}
Next we investigate the power of some of the methods from section \ref{sec:ReviewMTP} for the problem of testing the global hypothesis $H_0= H^1_0 \cap \cdots \cap H^K_0$, i.e. the probability of rejecting at least one hypothesis when at least one of the calibration hypotheses is false. Since we are interested only in the probability of rejecting at least one hypothesis, it suffices to consider only (Bonf) and (Hom) from group A and (d-Bonf) from group C as well as (BH) and (a-BH) from group B. We study the power of these procedures for the upgrade-downgrade and the dispersion setting introduced in section \ref{sec:ExampPFMisClass}. Additionally, we compare these results to the power of (HL) for detecting violation of $H_0$. For each combination of $s$ and $N$ the corresponding misclassification matrix was generated and $N_{sim}=10000$ simulations of default numbers $O_j$ for classes with $\widehat{n}_j \neq 0$ were carried out. This means that the standard error is bounded by $0.5\%$.
\paragraph{Results for upgrade-downgrade misclassification}
Figure \ref{fig:FWER_10000_UpDownGrade} depicts the simulated rejection probabilities in the case of alternatives of the upgrade-downgrade type for shifts $s= -5, \ldots, 5$ and for number of borrowers $N_{PF}=100, \ldots, 1000$.
\begin{center}
[Fig. \ref{fig:FWER_10000_UpDownGrade} about here]
\end{center} 
%\begin{sidewaysfigure}[h]
%	\centering
%		%\includegraphics[width=1.00\textwidth]{DistrTMIN.pdf}
%		\includegraphics[width=1.00\textwidth]{FWER_10000_UpDownGrade.pdf}
%	\caption{Probability of rejecting the overall hypothesis under upgrade-downgrade alternatives for sample sizes $100, \ldots,1000$. Hosmer-Lemeshow test ($\circ$), Bonferroni ($+$), Benjamini-Hochberg ($\triangledown$) and discrete Bonferroni ($\boxtimes$).}
%	\label{fig:FWER_10000_UpDownGrade}
%\end{sidewaysfigure}
%\begin{figure}[hbtp]
%	\centering
%		%\includegraphics[width=1.00\textwidth]{DistrTMIN.pdf}
%		\includegraphics[width=1.00\textwidth]{FWER_10000_UpDownGrade.pdf}
%	\caption{Simulated probabilities of rejecting the overall hypothesis under upgrade-downgrade alternatives for sample sizes $100, \ldots,1000$. (HL) ($\circ$), (Bonf) ($+$), (Hom) (???), (BH) ($\triangledown$) and (d-Bonf) ($\boxtimes$).}
%	\label{fig:FWER_10000_UpDownGrade}
%\end{figure}
It shows that in most constellations the procedures from groups A and B perform comparably. Again, (d-Bonf) seems to be the most powerful of the multiple testing procedures investigated here. It always outperforms the procedures from A and B. For positive values of $s$, all MTPs seem to be superior to (HL), for negative values of $s$ it is vice versa, with (d-Bonf) still performing relatively similar to (HL). 
\paragraph{Results for dispersed misclassification}
As in the case of upgrade-downgrade misclassification there seems to be little difference in the power of the procedures from group A and B, cf. figure \ref{fig:FWER_10000_Dispersion}. 
\begin{center}
[Fig. \ref{fig:FWER_10000_Dispersion} about here]
\end{center} 
Again, the (d-Bonf) method seems to outperform both other groups. For small sample sizes up to $N_{PF}=300$, the (HL) test outperforms all MTPs. For greater sample sizes, all MTPs seem to superior to (HL) for large values of the dispersion parameter.
%\begin{sidewaysfigure}[h]
%	\centering
%		%\includegraphics[width=1.00\textwidth]{FWER_10000_Dispersion.pdf}
%		\includegraphics[width=1.00\textwidth]{FWER_10000_Dispersion.pdf}
%	\caption{Probability of rejecting the overall hypothesis under dispersion alternatives for sample sizes $100, \ldots,1000$. Hosmer-Lemeshow test ($\circ$), Bonferroni ($+$), Benjamini-Hochberg ($\triangledown$) and discrete Bonferroni ($\boxtimes$).}
%	\label{fig:FWER_10000_Dispersion}
%\end{sidewaysfigure}
%\begin{figure}[bh]
%	\centering
%		%\includegraphics[width=1.00\textwidth]{FWER_10000_Dispersion.pdf}
%		\includegraphics[width=1.00\textwidth]{FWER_10000_Dispersion.pdf}
%	\caption{Simulated probabilities of rejecting the overall hypothesis under dispersion alternatives for sample sizes $100, \ldots,1000$. (HL) ($\circ$), (Bonf) ($+$), (Hom) (???), (BH) ($\triangledown$) and (d-Bonf) ($\boxtimes$).}
%	\label{fig:FWER_10000_Dispersion}
%\end{figure}

%% file: StandardPoors.tex
\section{Empirical study} \label{sec:EmpiricalStudy}
In this section we apply MTPs to empirical default data, using corporate data and PD estimates presented in Bl\"ochinger and Leippold (cf. \cite{BloeLeipp10} for more details). Table \ref{tab:PDEstimatesDurationCluster} presents two PD estimates for S\&P rating classes based on the duration and cohort approach. The estimates were obtained using training data from 1981 to 2002 and we follow Bl\"ochinger and Leippold and perform backtesting for the years 2003 to 2008. While they focus on the overall calibration resp. calibration concerning level and shape, our goal is again to identify which of the 17 rating classes are miscalibrated. 

Column $r$ in tables \ref{tab:AnalysisBLoechLeippCluster} and \ref{tab:AnalysisBLoechLeippDuration} lists which of the MTPs detected miscalibrated PDs. As in the simulation studies of section \ref{sec:SimStudy}, the procedures from group A are inferior to those from groups B and C. For both types of PD estimates, group B is able to identify some additional conspicuous PD estimates as compared to group C. Note also, that there are several classes, where (a-BH) performs strictly better than (BH). Hence group B outperforms group C for these validation samples. The MTP analysis could seem to suggest that except for the year 2008, miscalibration is mainly a feature of the rating classes with high PD. This conclusion may again be questionable in view of the low power for the high rating classes (see also the analysis in \ref{sssec:SimSinglePF}).  
\begin{center}
[Table \ref{tab:AnalysisBLoechLeippCluster} about here]
\end{center}
\begin{center}
[Table \ref{tab:AnalysisBLoechLeippDuration} about here]
\end{center}
If the same procedures are used to test the overall calibration hypothesis, then the findings are for the major part similar to the results described for the independence case in \cite{BloeLeipp10}: For the years 2004--2008 (HL), (BL) and all MTPs produce significant findings at the 5\% level. For 2003 none of the MTP, nor the (exact) HL test detect any miscalibration. Only the BL test is able to reject the calibration hypothesis for the PD estimates derived from the cluster approach. 

The BL test has the advantage that if the assumed two component model holds true, it may be possible to identify the component(s) that lead to rejection of the calibration hypothesis. Note however that this need not always be the case as the data for the cluster PD estimates in 2003 illustrates. Here both the level and shape components are insignificant (both $p$-values equal $0.2061$) while the global test is significant ($p$-value $0.0243$). Approaches based on MTPs on the other hand are more non parametric in nature, i.e. no parametric model in the sense of the two component model in \cite{BloeLeipp10} is assumed. Since MTPs provide a per-class assessment, this means that they may give more detailed information than the BL test. The BL test on the other hand has the appeal of providing results that can be interpreted in terms of level and shape within the assumed model.

%% file: Extension.tex
\section{Extension to dependent defaults} \label{sec:ExtensionDependentDefaults}
In this section we sketch, how the \MinP methods used in this paper for independent defaults can be extended to dependent defaults. A more detailed description and analysis will be the subject of future work.

In the one-factor model, which is also used in the IRB-approach of Basel II, the credit worthiness of each borrower $i$ is modeled as
\begin{align*}
A_i &=\sqrt{\rho} \cdot Z + \sqrt{1-\rho} \cdot U_i
\end{align*}
with $0<\rho<1$ where $Z, U_1,\ldots,U_N \sim \NormVert(0,1)$ iid, cf. e.g. \cite{Tasche05}. The risk factor $Z$ denotes the systematic risk component which is common to all borrowers and $U_i$ is the idiosyncratic risk that is specific to borrower $i$. The 'asset correlation' $\rho$ describes the dependency of individual defaults on the systematic risk component. Accordingly, default indicators $X_{ij}$ for borrower $i$ in rating class $j$ can be defined by
\begin{align*}
X_{ij} &=
\begin{cases}
1 \qquad & A_i \le \Phi^{-1}(p_j)\\
0 \qquad & \text{else},
\end{cases}
\end{align*}
where $\Phi^{-1}$ is the quantile function of $\NormVert(0,1)$. Note that the $X_{ij}\sim \BinVert(1,p_j)$ but they are no longer independent. Assume that the test statistics $T_j$ resp. $p$ values $PV_j$ per class are measurable functions of $S_j=(X_{1j},\ldots,X_{\widehat{n}_j j})$. Then the \MinP approach can be implemented as follows:
\begin{enumerate}
	\item Obtain the $p$ value functions $PV_j=PV(s_j)=PV_j(x_{1j},\ldots,x_{\widehat{n}_j j})$.
	\item Obtain the distribution function of the $PV_j$'s. 
	\item Obtain the distribution function $F_{\{1, \ldots,K\}}$ of $\min (PV_1,\ldots,PV_K)$.
	\item For a given sample of defaults $(s_1,\ldots,s_K)$ calculate marginal $p$ values
\begin{align*}
pv_1 &= PV(s_1)=PV_1(x_{11},\ldots,x_{\widehat{n}_1 1})\\
 &\vdots \\
pv_K &=  PV(s_K)=PV_K(x_{1K},\ldots,x_{\widehat{n}_K K})
\intertext{resp. adjusted $p$ values}
pv_1' &= F_{\{1, \ldots,K\}}(pv_1)\\
 &\vdots \\
pv_K' &= F_{\{1, \ldots,K\}}(pv_K)
\end{align*}
and continue in the spirit of section \ref{ssec:The Min(P)-approach for discrete distributions}.
\end{enumerate}
In steps 1 through 3 it may not be feasible to obtain exact quantities. In these cases, simulation can be used to obtain sufficiently accurate estimates.

Since by assumption, $PV_j$ is a function of $S_j=(X_{1j},\ldots,X_{\widehat{n}_j j})$ and the distribution of this vector depends only on $p_j$ (and the fixed asset correlation), proposition \ref{prop:SuffCondSPC} shows that (SPC) is fulfilled and therefore the procedure sketched above also maintains control of the FWER.

%% file: Discussion.tex
\section{Discussion} \label{sec:Discussion}
In this paper we have applied MTPs to testing the calibration of PD estimates in credit rating systems with a view towards identifying miscalibrated PD estimates. We have considered procedures that control FWER and FDR and have investigated their performance in a simulation setting and for empirical data.

For FWER, the results show that the power of 'standard' procedures can be substantially improved by \MinP methods, which take the discreteness of data into account. These methods also perform well as tests of the overall calibration hypothesis. In addition, we have used the more explorative FDR methodology for identifying conspicuous PD estimates. In the simulation study, the power of these methods was roughly comparable to the \MinP methods, while for the empirical data they outperformed the \MinP methods. This may be due to the higher number of rating grades resp. tests performed. Note also that no attempt was made to adapt the FDR procedures to the discreteness of the data. If this were done in an appropriate way, their power might be considerably enhanced. In this sense the presentation given in this paper is somewhat biased against the FDR approach. Altogether, we conclude that in the framework of independent defaults considered here, MTPs can serve as helpful tools for identifying miscalibrated resp. conspicuous PD estimates.

We have also seen that MTPs frequently perform well as tests for the global calibration hypothesis. The question whether to use these methods or a global test has been discussed in the statistical literature. Westfall and Wolfinger sum up the situation as follows (see \cite{WestWolf1997}):
\begin{quote}
	\textit{'The global test will have higher power in situations where there is a mild departure from the null for many tests, whereas our [\MinP] methods have high power, when there are marked departures from the null at only a few sites [tests]. Regardless of power comparisons, a major problem with global tests is their failure to isolate specific significances.' }
\end{quote}

While we have concentrated on the independence case for the sake of illustrating the main ideas as simply as possible, credit default events are usually not independent and procedures that aim to identify miscalibrated PD estimates should take this into account. One possible approach would consist of using the Bonferroni variant of the \MinP method resp. the Benjamini-Yektutieli variant of the (BH) approach as 'worst case' types of dependency. However, this may be overly conservative. It seems more promising to extend the \MinP method as described in section \ref{sec:ExtensionDependentDefaults} in order to account for specific forms of dependencies. This will be the subject of future work.

\subsection*{Acknowledgements} The author would like to thank Marcus R.W. Martin for helpful comments.

%To control the rate of false rejections we have restricted ourselves to FWER and FDR controlling procedures, but it could make sense to investigate other measures as well. Consider, for instance, the $k$-FWER, i.e. the probability of erroneously rejecting not one or more, but rather $k$ or more hypotheses, see \cite{Sarkar2007} for this and other generalisations. This could be used e.g. as the basis of a traffic lights approach in the sense that as long as less than $k$ PD estimates are identified the overall calibration is not questioned.
%
%      
%\begin{itemize}
%	\item incorporating dependencies
%	\item traffic light approach (Henking)
%	\item MTP approach 'automatically' corrects for the number of rating categories used (but power problem)
%\end{itemize}

%% file: Tables.tex
\section*{Tables}

\begin{table}[htb]
	\centering
		\begin{tabular}{lrrrrrrr}\hline\hline
class $i$ &1 & 2 & 3 & 4 & 5 & 6 & 7\\
$100 \cdot p_i$& 0.015 & 0.03 & 0.06 & 0.11 & 0.2 & 0.35 & 0.6\\
$q_i$& 0.009 & 0.014 & 0.053 & 0.070 & 0.133 & 0.133 & 0.164\\\hline
&&&&&&&\\\hline\hline
class $i$ &8 & 9 & 10 & 11 & 12 & 13 & 14\\
$100 \cdot p_i$ & 1.05 & 1.85 & 3.25 & 5.7 & 10.0 & 17.5 & 33.8\\
$q_i$& 0.149 & 0.123 & 0.077 & 0.046 & 0.020 & 0.008 & 0.002\\\hline
\end{tabular}  
\caption{True PDs and probability distribution of borrowers to (true) rating classes in the simulation study}
\label{tab:TableSimStudy}
\end{table}

\begin{table}[hbt]
	\centering
		\begin{tabular}{lrr}\hline\hline
S\&P rating & PD (Duration) & PD (Cluster)  \\\hline

       AAA &       0.02 &          1.00 \\

       AA+ &       0.05 &          1.00 \\

        AA &       0.43 &          1.00 \\

       AA- &       0.44 &       3.84 \\

        A+ &       0.46 &        5.20 \\

         A &       0.84 &       6.49 \\

        A- &          1.00 &       6.49 \\

      BBB+ &       4.67 &      31.37 \\

       BBB &      11.65 &      36.23 \\

      BBB- &      14.53 &      40.12 \\

       BB+ &      33.01 &      55.01 \\

        BB &      45.64 &     116.33 \\

       BB- &      88.51 &     207.18 \\

        B+ &     175.41 &      349.80 \\

         B &     758.33 &     982.01 \\

        B- &     1343.30 &    1430.16 \\

       CCC &    4249.04 &    2853.54 \\\hline
\end{tabular}  
\caption{Estimated probabilities of default (PD in bps, i.e. 1/100\%) for the duration and cluster approaches for the S\&P data from \cite{BloeLeipp10}}
\label{tab:PDEstimatesDurationCluster}
\end{table}

{\small
\begin{sidewaystable}[!tbp]
 \begin{center}
 % Table generated by Excel2LaTeX from sheet 'Kopie von DetailsAnalysisBLoech'
\begin{tabular}{lrrrrrrrrrrrrrrrrrr}\hline\hline

           &            \multicolumn{ 3}{c}{2003} &            \multicolumn{ 3}{c}{2004} &            \multicolumn{ 3}{c}{2005} &            \multicolumn{ 3}{c}{2006} &            \multicolumn{ 3}{c}{2007} &            \multicolumn{ 3}{c}{2008} \\ 
 \cmidrule(lr){2-4}  \cmidrule(lr){5-7} \cmidrule(lr){8-10} \cmidrule(lr){11-13} \cmidrule(lr){14-16} \cmidrule(lr){17-19}
           &         $n$ &         $d$ &         $r$ &         $n$ &         $d$ &         $r$ &         $n$ &         $d$ &         $r$ &         $n$ &         $d$ &         $r$ &         $n$ &         $d$ &         $r$ &         $n$ &         $d$ &         $r$ \\\hline

       AAA &        103 &          0 &            &         94 &          0 &            &         93 &          0 &  &         86 &          0 &            &         90 &          0 &            &         93 &          0 &            \\

       AA+ &         52 &          0 &            &          4 &          0 &            &         41 &          0 &  &         41 &          0 &            &         39 &          0 &            &         50 &          0 &            \\

        AA &        164 &          0 &            &        157 &          0 &            &        143 &          0 &  &        152 &          0 &            &        168 &          0 &            &        223 &          1 &            \\

       AA- &        221 &          0 &            &        206 &          0 &            &        217 &          0 &  &        221 &          0 &            &        261 &          0 &            &        235 &          1 &            \\

        A+ &        335 &          0 &            &        324 &          0 &            &        321 &          0 &  &        319 &          0 &            &        296 &          0 &            &        302 &          1 &            \\

         A &        386 &          0 &            &        413 &          1 &            &        417 &          0 &  &        450 &          0 &            &        450 &          0 &            &        453 &          1 &            \\

        A- &        404 &          0 &            &        427 &          0 &            &        475 &          0 &  &        523 &          0 &            &        492 &          0 &            &        500 &          3 &  4--8 \\

      BBB+ &        415 &          0 &            &        447 &          0 &            &        480 &          0 &  &        511 &          0 &            &        494 &          0 &            &        507 &          1 &            \\

       BBB &        492 &          1 &            &        545 &          0 &            &        557 &          1 &            &        513 &          0 &            &        513 &          0 &            &        485 &          3 &            \\

      BBB- &        360 &          2 &            &        390 &          0 &            &        384 &          0 &            &        266 &          0 &            &        349 &          0 &            &        396 &          3 &            \\

       BB+ &        191 &          1 &            &        235 &          0 &            &        246 &          1 &            &        237 &          1 &            &        254 &          0 &            &        252 &          3 &            \\

        BB &        292 &          3 &            &        288 &          2 &            &        290 &          0 &           &        290 &          0 &            &        285 &          1 &            &        296 &          2 &            \\

       BB- &        322 &          1 &            &        351 &          3 &            &        362 &          1 &  4,5,8 &        354 &          2 &            &        395 &          1 &  4--8 &        425 &          3 &            \\

        B+ &        358 &          7 &            &        384 &          2 & 1--8 &        445 &          4 & 1--8 &        494 &          3 & 1--8 &        456 &          1 & 1--8 &        456 &         15 &            \\

         B &        181 &         11 &            &        223 &          7 & 1--8 &        262 &          8 & 1--8 &        324 &          3 & 1--8 &        426 &          0 & 1--8 &        571 &         21 & 1--8 \\

        B- &        116 &         12 &            &        120 &          4 & 1--8 &        140 &          5 & 1--8 &        163 &          3 & 1--8 &        181 &          2 & 1--8 &        205 &         17 &          5 \\

       CCC &        149 &         55 &            &        118 &         21 &        4,5 &        102 &         11 & 1--8 &         90 &         14 &  4--8 &         86 &         16 &            &         78 &         26 &            \\\hline
$p$ values (HL) &\multicolumn{ 3}{c}{0.2012} &            \multicolumn{ 3}{c}{0.0336} &            \multicolumn{ 3}{c}{0.0275} &            \multicolumn{ 3}{c}{0.0159} &            \multicolumn{ 3}{c}{0.0128} &            \multicolumn{ 3}{c}{0.0082} \\ 
$p$ values (BL) &\multicolumn{ 3}{c}{0.0243} &            \multicolumn{ 3}{c}{0.0000} &            \multicolumn{ 3}{c}{0.0000} &            \multicolumn{ 3}{c}{0.0000} &            \multicolumn{ 3}{c}{0.0000} &            \multicolumn{ 3}{c}{0.0000} \\ \hline
\end{tabular}  
\end{center}
\caption{Identification of miscalibrated rating classes for the S\&P data, using PD estimates from the cluster approach. For each year, column $n$ presents the number of borrowers, column $d$ lists the number of observed defaults, and column $r$ lists which methods produced significant findings after correcting for multiplicity (Bonf (1), Hol (2), Hom (3), BH (4), a-BH (5), d-Bonf (6), d-Ind (7), sd-d-Bonf (8))}
\label{tab:AnalysisBLoechLeippCluster}
\end{sidewaystable}
}

\begin{sidewaystable}[!tbp]
 \begin{center}
% Table generated by Excel2LaTeX from sheet 'DetailsAnalysisBLoechLeippDurat'
\begin{tabular}{lrrrrrrrrrrrrrrrrrr}\hline\hline

           &            \multicolumn{ 3}{c}{2003} &            \multicolumn{ 3}{c}{2004} &            \multicolumn{ 3}{c}{2005} &            \multicolumn{ 3}{c}{2006} &            \multicolumn{ 3}{c}{2007} &            \multicolumn{ 3}{c}{2008} \\
\cmidrule(lr){2-4}  \cmidrule(lr){5-7} \cmidrule(lr){8-10} \cmidrule(lr){11-13} \cmidrule(lr){14-16} \cmidrule(lr){17-19}
           &     $n$ &     $d$ &       $r$ &     $n$ &     $d$ &       $r$ &     $n$ &     $d$ &       $r$ &     $n$ &     $d$ &       $r$ &     $n$ &     $d$ &       $r$ &     $n$ &     $d$ &       $r$ \\\hline

       AAA &        103 &          0 &            &         94 &          0 &            &         93 &          0 &            &         86 &          0 &            &         90 &          0 &            &         93 &          0 &            \\

       AA+ &         52 &          0 &            &         41 &          0 &            &         41 &          0 &            &         41 &          0 &            &         39 &          0 &            &         50 &          0 &            \\

        AA &        164 &          0 &            &        157 &          0 &            &        143 &          0 &            &        152 &          0 &            &        168 &          0 &            &        223 &          1 &      4,5,8 \\

       AA- &        221 &          0 &            &        206 &          0 &            &        217 &          0 &            &        221 &          0 &            &        261 &          0 &            &        235 &          1 &      4,5 \\

        A+ &        335 &          0 &            &        324 &          0 &            &        321 &          0 &            &        319 &          0 &            &        296 &          0 &            &        302 &          1 &        4,5 \\

         A &        386 &          0 &            &        413 &          1 &            &        417 &          0 &            &        450 &          0 &            &        450 &          0 &            &        453 &          1 &          5 \\

        A- &        404 &          0 &            &        427 &          0 &            &        475 &          0 &            &        523 &          0 &            &        492 &          0 &            &        500 &          3 & 1--8 \\

      BBB+ &        415 &          0 &            &        447 &          0 &            &        480 &          0 &            &        511 &          0 &            &        495 &          0 &            &        507 &          1 &            \\

       BBB &        492 &          1 &            &        545 &          0 &            &        557 &          1 &            &        513 &          0 &            &        513 &          0 &            &        485 &          3 &        4,5 \\

      BBB- &        360 &          2 &            &        390 &          0 &            &        384 &          0 &            &        366 &          0 &            &        349 &          0 &            &        396 &          3 &        4,5 \\

       BB+ &        191 &          1 &            &        235 &          0 &            &        246 &          1 &            &        237 &          1 &            &        254 &          0 &            &        252 &          3 &          5 \\

        BB &        292 &          3 &            &        288 &          2 &            &        290 &          0 &            &        290 &          0 &            &        285 &          1 &            &        296 &          2 &            \\

       BB- &        322 &          1 &            &        351 &          3 &            &        362 &          1 &            &        354 &          2 &            &        395 &          1 &            &        425 &          3 &            \\

        B+ &        358 &          7 &            &        384 &          2 &            &        445 &          4 &            &        494 &          3 &            &        456 &          1 &  4--8 &        456 &         15 &        4,5 \\

         B &        181 &         11 &            &        223 &          7 &  4,5,7,8 &        262 &          8 & 2--8 &        324 &          3 & 1--8 &        426 &          0 & 1--8 &        571 &         21 & 1--8 \\

        B- &        116 &         12 &            &        120 &          4 & 1--8 &        140 &          5 & 1--8 &        163 &          3 & 1--8 &        181 &          2 & 1--8 &        205 &         17 &          5 \\

       CCC &        149 &         55 &            &        118 &         21 & 1--8 &        102 &         11 & 1--8 &         90 &         14 & 1--8 &         86 &         16 & 1--8 &         78 &         26 &            \\\hline
$p$ values (HL)      &            \multicolumn{ 3}{c}{0.3500} &            \multicolumn{ 3}{c}{0.0213} &            \multicolumn{ 3}{c}{0.0318} &            \multicolumn{ 3}{c}{0.0289} &            \multicolumn{ 3}{c}{0.0219} &            \multicolumn{ 3}{c}{0.0005} \\
$p$ values (BL) &\multicolumn{ 3}{c}{0.1873} &            \multicolumn{ 3}{c}{0.0000} &            \multicolumn{ 3}{c}{0.0000} &            \multicolumn{ 3}{c}{0.0000} &            \multicolumn{ 3}{c}{0.0000} &            \multicolumn{ 3}{c}{0.0000} \\ \hline
\end{tabular}  

\end{center}
\caption{Identification of miscalibrated rating classes for the S\&P data, using PD estimates from the cluster approach. For each year, column $n$ presents the number of borrowers, column $d$ lists the number of observed defaults, and column $r$ lists which methods produced significant findings after correcting for multiplicity (Bonf (1), Hol (2), Hom (3), BH (4), a-BH (5), d-Bonf (6), d-Ind (7), sd-d-Bonf (8))}
\label{tab:AnalysisBLoechLeippDuration}
\end{sidewaystable}

%% file: Figures.tex
\section*{Figures}
\begin{figure}[htb]
	\centering
		\includegraphics[height=0.31\textheight]{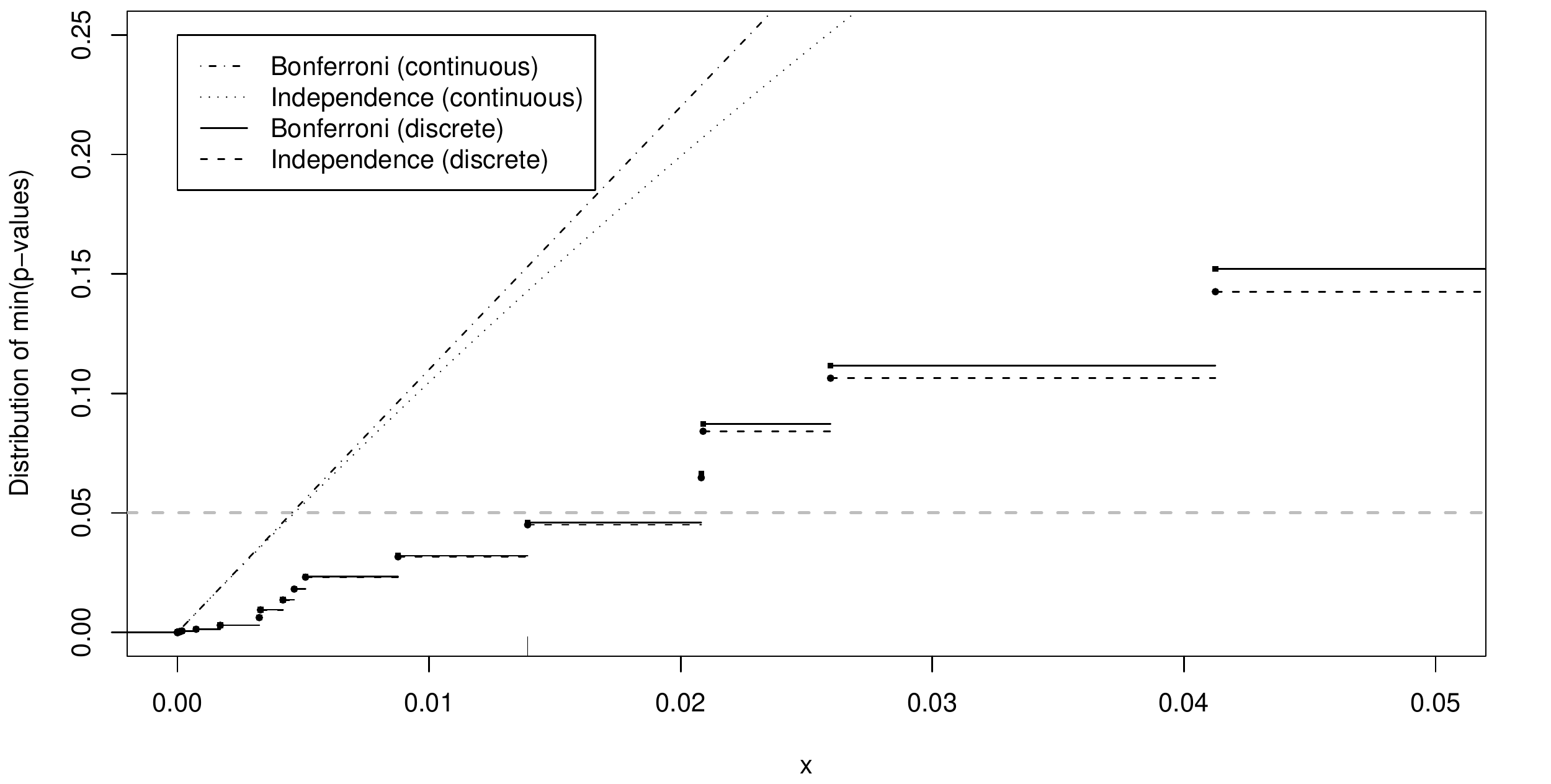}
	\caption{Distribution functions of $F^{Ind}_{\{1,\ldots,11\}}$ and $F^{Bonf}_{\{1,\ldots,11\}}$ in the continuous and the discrete case}
	\label{fig:DistrTMIN}
\end{figure}
%\begin{figure}[h]
%	\centering
%		%\includegraphics[width=1.00\textwidth]{PortfolioGraph.pdf}
%		\includegraphics[height=0.30\textheight]{PortfolioGraph.pdf}
%	\caption{Distribution of rating classes for example portfolio (PDs are given below class numbers)}
%	\label{fig:PortfolioGraph}
%\end{figure} 
\begin{figure}[hbt]
	\centering
		\includegraphics[height=0.31\textheight]{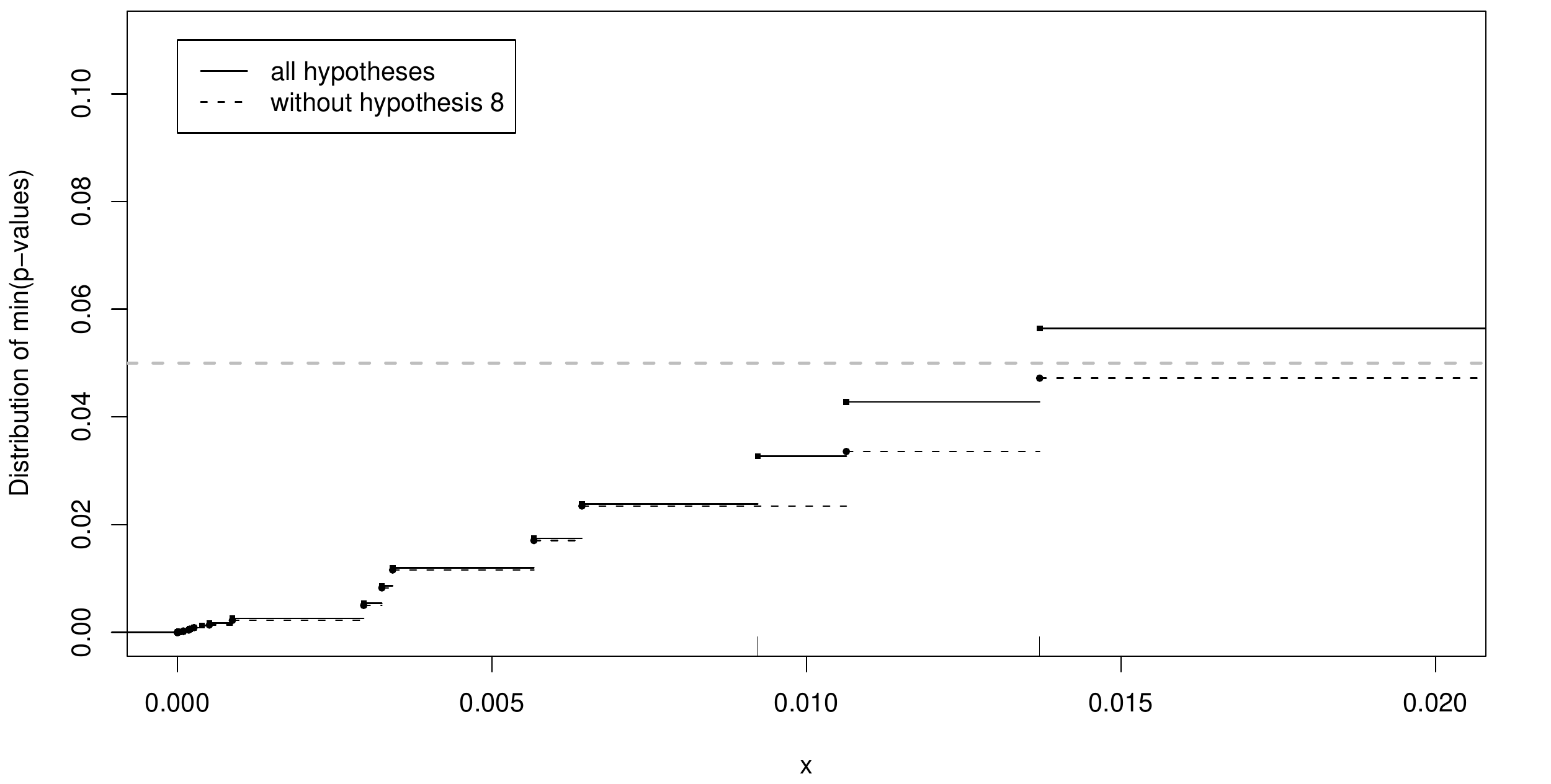}
	\caption{Distribution functions for \MinP for the first and second step in (sd-d-Bonf)}
	\label{fig:ExampMTPstepdown}
\end{figure} 
\begin{figure}[htbp]
	\centering
		\includegraphics[height=0.33\textheight]{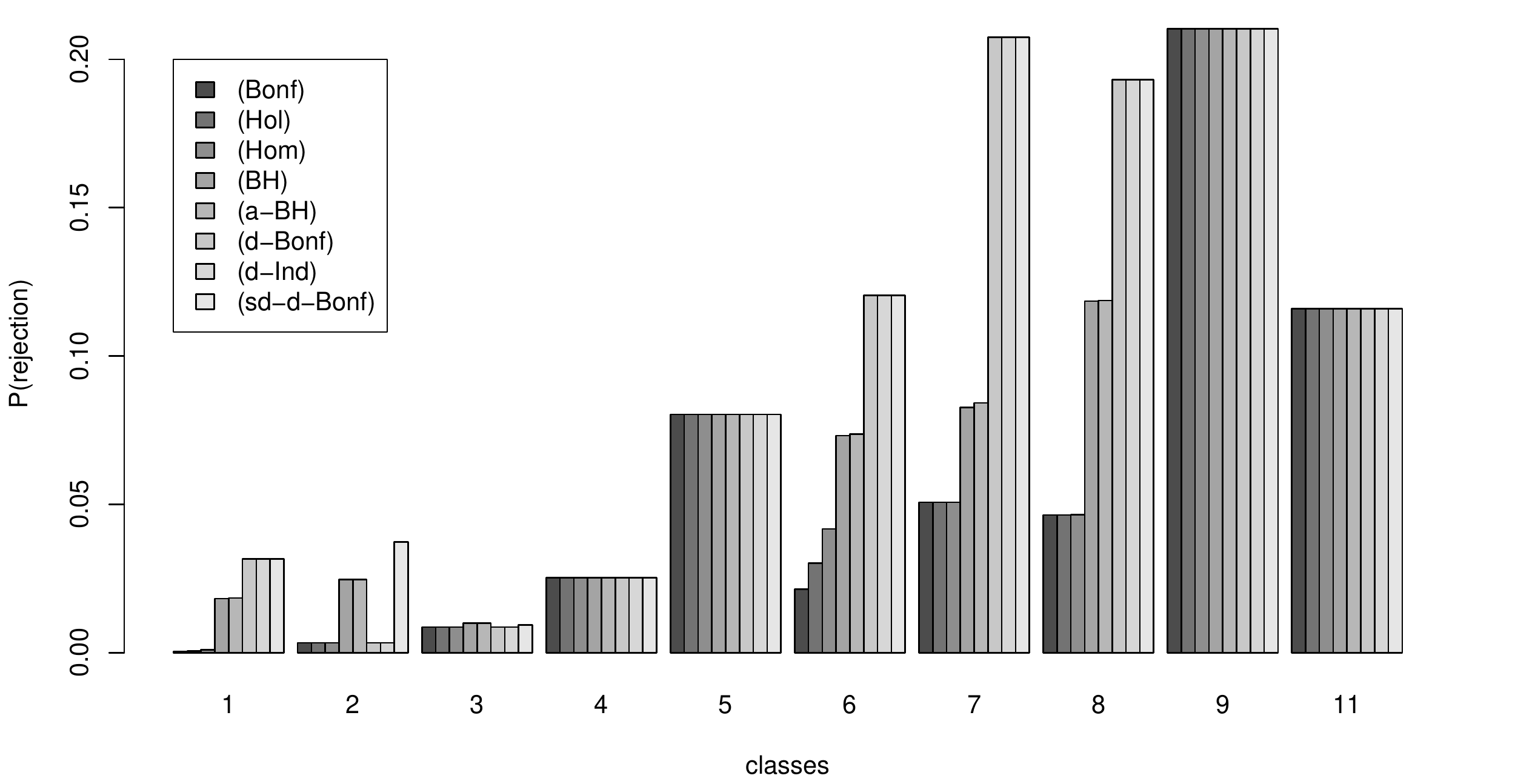}
	\caption{Simulated probabilities of rejecting null hypotheses corresponding to the rating classes for $N=300$ and upgrade-downgrade alternative with $s=-3$}
	\label{fig:Identify_300_-3_100}
\end{figure} 
\begin{sidewaysfigure}[htbp]
	\centering
		\includegraphics[width=1\textwidth]{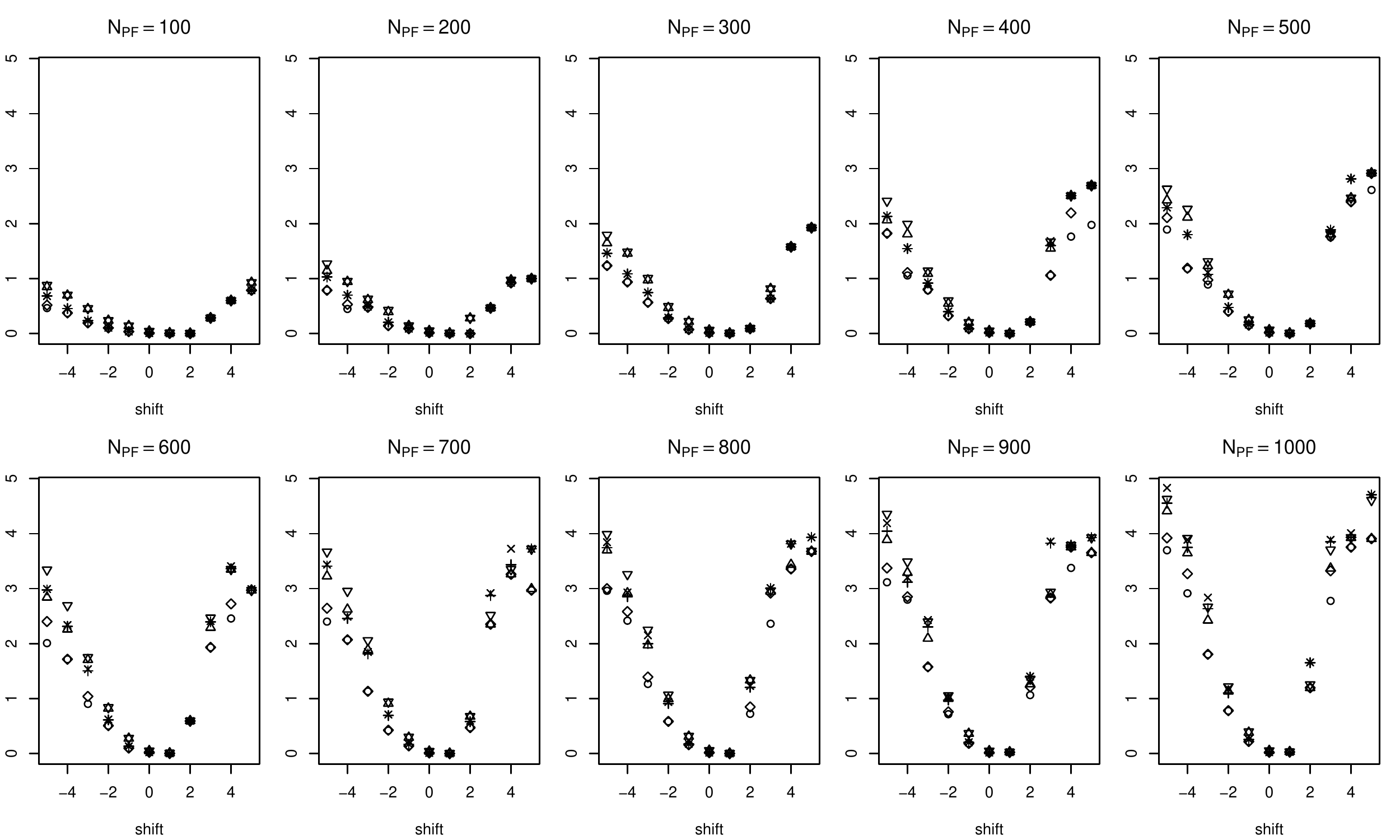}
	\caption{Average number of rejected hypotheses under upgrade-downgrade alternatives for sample sizes $100, \ldots,1000$. (Bonf) ($\circ$), (Hom) ($\diamondsuit$), (BH) ($+$), (a-BH) ($\times$), (d-Bonf) ($\triangle$) and (sd-d-Bonf) ($\triangledown$).}
	\label{fig:Identify_Ave_num_rejections_UpDownGrade}
\end{sidewaysfigure} 
\begin{sidewaysfigure}[htbp]
	\centering
		\includegraphics[width=1\textwidth]{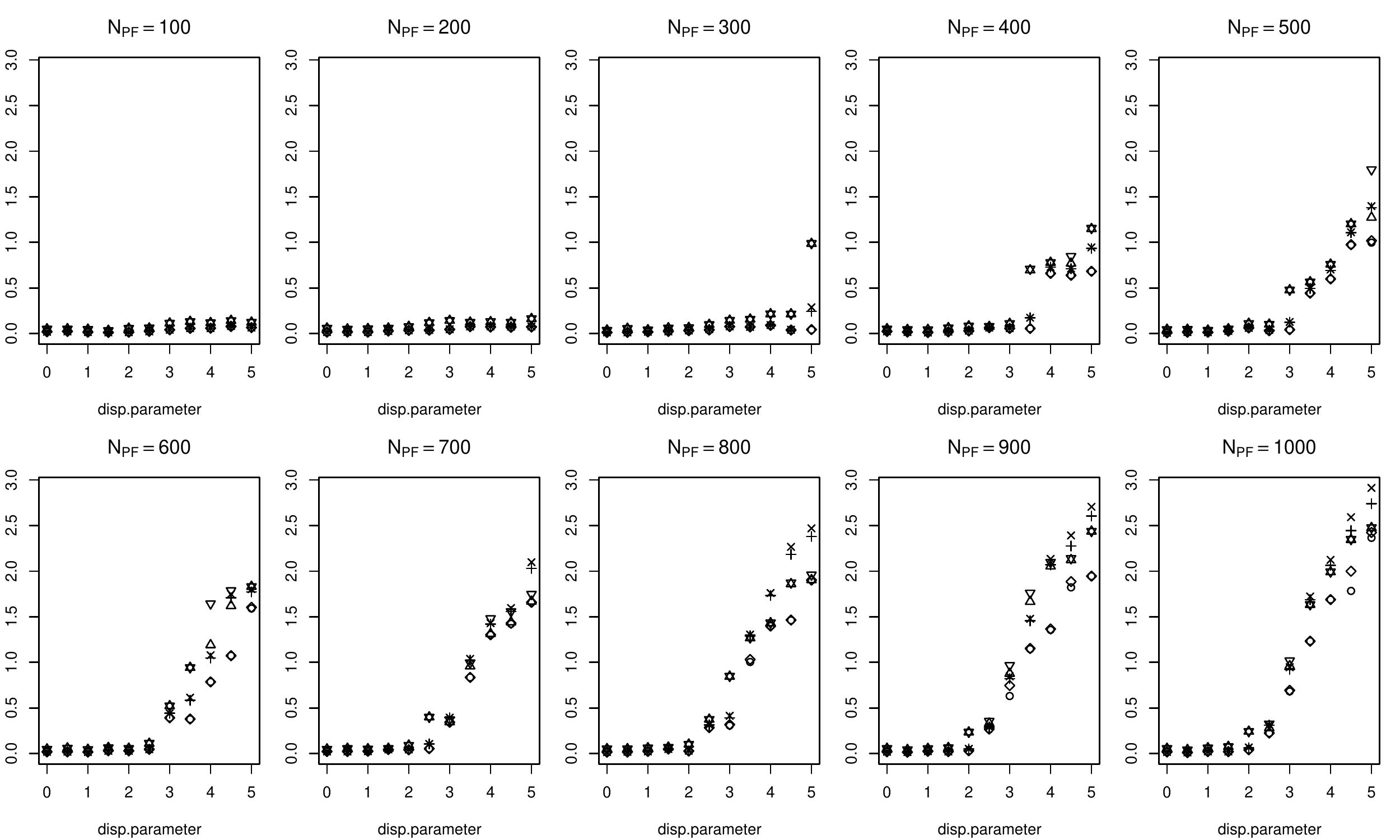}
	\caption{Average number of rejected hypotheses under dispersion alternatives for sample sizes $100, \ldots,1000$. (Bonf) ($\circ$), (Hom) ($\diamondsuit$), (BH) ($+$), (a-BH) ($\times$), (d-Bonf) ($\triangle$) and (sd-d-Bonf) ($\triangledown$).}
	\label{fig:Identify_Ave_num_rejections_Dispersion}
\end{sidewaysfigure} 
\begin{sidewaysfigure}[htbp]
	\centering
		\includegraphics[width=1.00\textwidth]{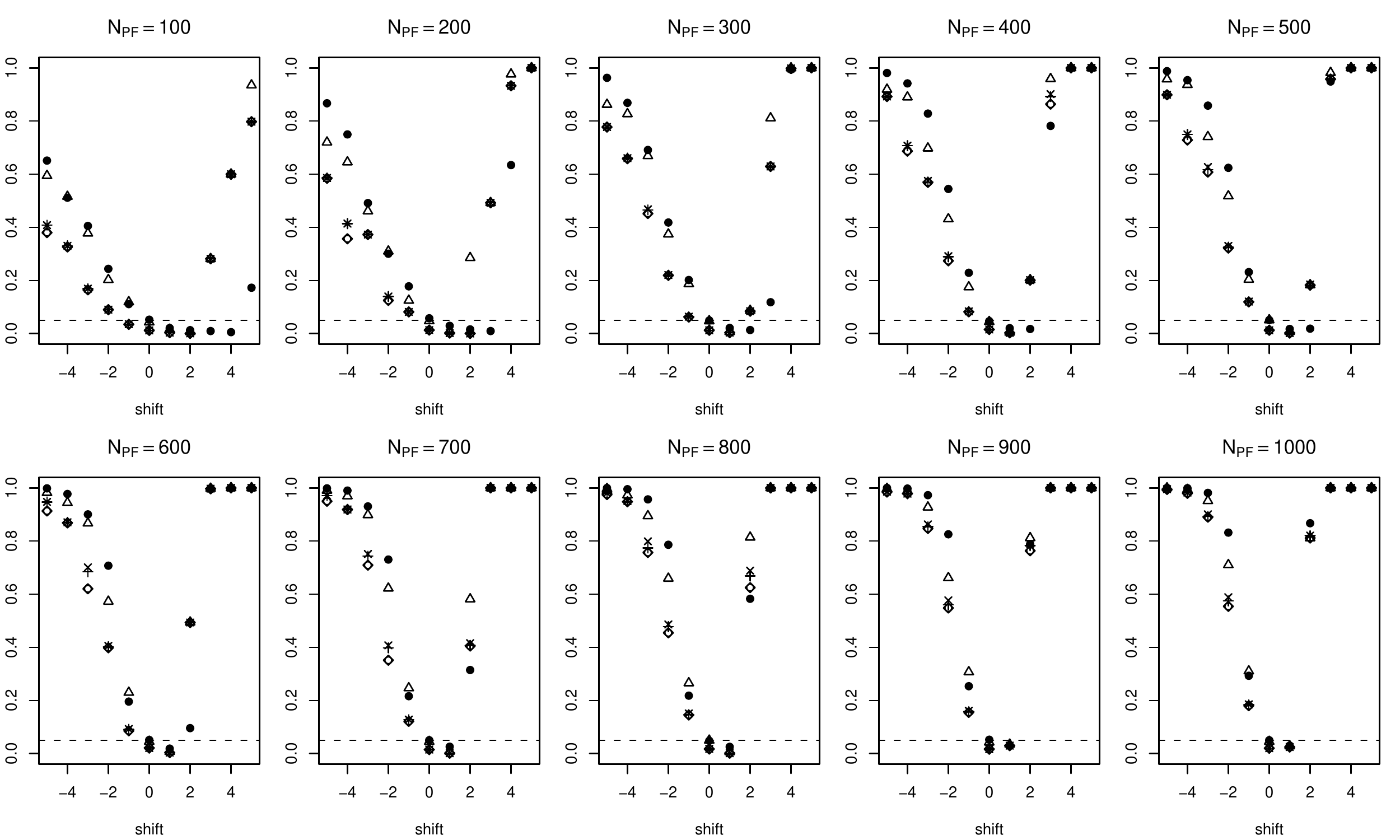}
	\caption{Simulated probabilities of rejecting the overall hypothesis under upgrade-downgrade alternatives for sample sizes $100, \ldots,1000$. (HL) ($\bullet$), (Bonf) ($\circ$), (Hom) ($\diamondsuit$), (BH) ($+$), (a-BH) ($\times$), (d-Bonf) ($\triangle$).}
	\label{fig:FWER_10000_UpDownGrade}
\end{sidewaysfigure}
\begin{sidewaysfigure}[htbp]
	\centering
		\includegraphics[width=1.00\textwidth]{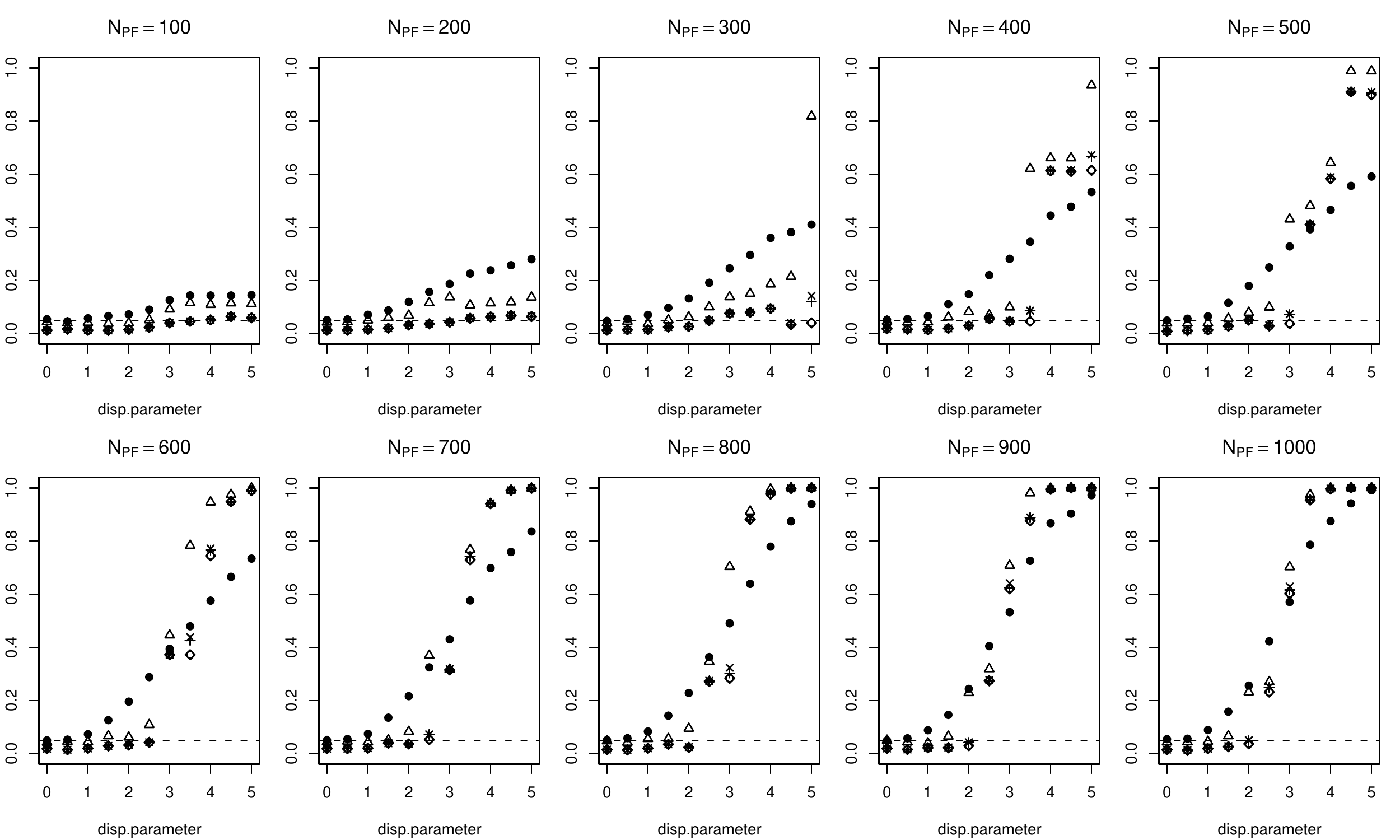}
	\caption{Simulated probabilities of rejecting the overall hypothesis under dispersion alternatives for sample sizes $100, \ldots,1000$. (HL) ($\bullet$), (Bonf) ($\circ$), (Hom) ($\diamondsuit$), (BH) ($+$), (a-BH) ($\times$), (d-Bonf) ($\triangle$).}
	\label{fig:FWER_10000_Dispersion}
\end{sidewaysfigure}
%\begin{figure}[htbp]
%	\centering
%		\includegraphics[height=0.33\textheight]{AnalysisBLoechLeippDuration.pdf}
%		%\includegraphics[width=1.00\textwidth]{AnalysisBLoechLeippDuration.pdf}
%	\caption{Number of miscalibrated classes for the duration approach detected at overall significance level 5\% for the S\&P data}
%	\label{fig:AnalysisBLoechLeippDuration}
%\end{figure}
%\begin{figure}[htbp]
%	\centering
%		\includegraphics[height=0.33\textheight]{AnalysisBLoechLeippCluster.pdf}
%		%\includegraphics[width=1.00\textwidth]{AnalysisBLoechLeippCluster.pdf}
%	\caption{Number of miscalibrated classes for the cluster approach detected at overall significance level 5\% for the S\&P data}
%	\label{fig:AnalysisBLoechLeippCluster}
%\end{figure}